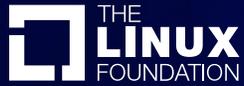
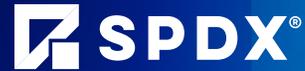

# Implementing AI Bill of Materials (AI BOM) with SPDX 3.0

A Comprehensive Guide to Creating AI and Dataset Bill of Materials.

**Authors**
Karen Bennet, Gopi Krishnan Rajbahadur, Arthit Suriyawongkul and Kate Stewart

October 2024

# Contents

# Overview

Artificial intelligence (AI) applications, especially those integrating open-source software and freely available data, are crucial for driving innovation, fostering collaboration, ensuring transparency, and democratizing access to AI technology. These applications significantly contribute to societal advancement and the broader adoption of responsible AI practices. However, they often incorporate numerous third-party components that can be designed for different environments, harbor security vulnerabilities, contain compromised data, or have unclear licensing terms.

In addition to traditional software risks, AI and data-intensive applications face unique challenges such as data security breaches arising from the misuse of data, including violations of consent and intent, and AI security threats like model tampering and adversarial attacks. Regulatory obligations for documenting non-traditional software engineering aspects, such as design decisions, known bias, energy consumption, and fundamental rights impact assessments, can significantly complicate AI risk management. Furthermore, the growing prevalence of open licenses for datasets and models, often with complex behavioral-use clauses, introduces new considerations for managing AI risks.

In this context, an AI bill of materials (AI BOM), which expands the traditional scope of software bill of materials (SBOM), is becoming increasingly crucial for managing increasingly complex systems and the broader adoption of AI. Capturing all relevant design decisions and development dependencies in AI system development is now considered a best practice and is often mandated by regulatory bodies and industry standards. An AI BOM enables organizations to systematically identify and track these decisions and dependencies, allowing them to proactively identify, manage, and mitigate risks related to technical security, functional safety, intellectual property, and regulatory compliance. As a result, AI applications can remain secure and resilient against potential threats.

The System Package Data Exchange specification version 3.0 (SPDX 3.0) aims to provide a machine-readable inventory of system components and documentation. The AI BOM within SPDX 3.0 incorporates relevant SPDX profiles to comprehensively document algorithms, data collection methods, libraries, frameworks, licensing information, standard compliance, security measures, and other tools involved in the development, testing, and deployment of AI applications. Additionally, while SPDX as a project does not take a stance on the necessity of licenses and the applicability of copyright laws in the context of AI and datasets, SPDX's metadata fields enable bill of materials producers and consumers to capture and communicate licensing-related information as stated by AI and dataset providers.

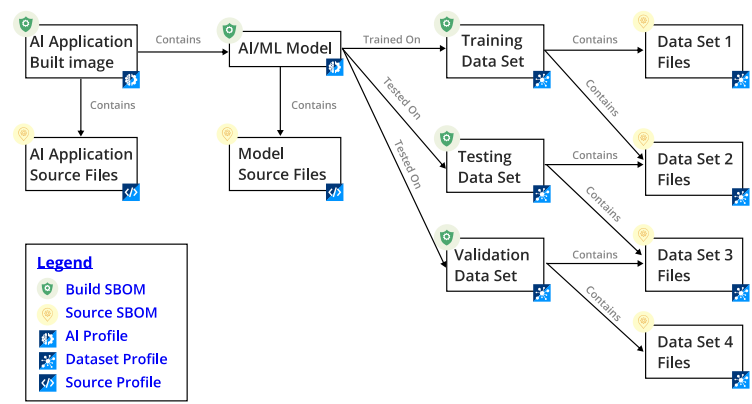

# Why SPDX Matters: Importance and Consequences

SPDX AI and Dataset Profiles serve as essential building blocks to promote transparency and accountability in the supply chain, while enabling the tracking of vulnerabilities, potential harms and risks. By providing a standardized documentation on AI systems' design, capabilities, and constraints, SPDX metadata for the AI Profile fosters trust and understanding among users, empowering them to make informed and safe decisions regarding their systems. In other words, these profiles not only enhance the reliability and traceability of AI systems but also contribute to their responsible use.

These documents are essential for developers, users, lawyers, regulators, and ethicists alike, as they foster a collaborative environment that addresses and manages the social implications of AI. Created and maintained by the organizations that developed the models, these profiles play a vital role in ensuring that AI technologies meet ethical standards and legal requirements. By providing standardized documentation, they facilitate responsible research and deployment within the AI ecosystem, promoting transparency and accountability throughout the supply chain.

SPFX 3.0 has been rearchitected to create profiles for the different information. We are focusing on AI and Dataset Profiles in this paper.

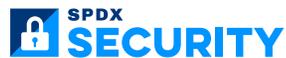 Security information - vulnerability details related to software

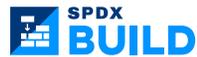 Build related information - provenance and reproducible builds

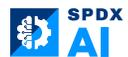 Information about AI models - ethical, security, and model data

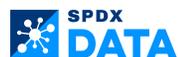 Information about datasets - AI and other data use cases

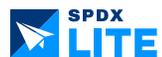 Minimal subset to support industry supply chain workflows

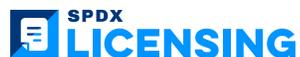 Information about copyrights and licenses - supports compliance

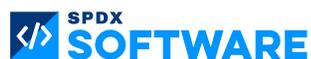 Information specific to software

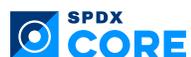 Information used across all profiles

# Motivation for Separating AI BOM into AI and Dataset Profiles

In SPDX (System Package Data Exchange), the AI and Dataset Profiles were separated to clarify the provenance and training methods used in creating an AI model.

The AI Profile is considered to include the machine learning (ML) use cases as well. The identification and categorization of software components in an AI/ML model is influenced by the datasets used to train and update a model. There is not always a direct relationship between a specific dataset and a specific model, as each can be acquired separately. AI Profile may refer to software components that are directly related to artificial intelligence functionalities, such as traditional machine learning algorithms, neural networks, large language models (LLMs) and other pre-processing and evaluation components. These components may have specific licensing considerations, usage restrictions, or security implications that are distinct from other types of software components.

Different from the AI Profile, the Dataset Profile represents components that primarily deal with data processing, storage, or management. These include databases and files that are composed of processed data. Separating these profiles helps in better understanding the data-related aspects of the software and its potential impact on privacy, compliance, or security.

By separating AI and Dataset Profiles, an AI BOM composed with SPDX aims to provide a more granular and comprehensive view of the software composition, making it easier for stakeholders to assess and manage the associated risks, obligations, and dependencies. However, it is important for the readers to note that an AI BOM is not just composed of fields from AI and Dataset Profiles (unless the objective is to only describe the model and/or dataset(s)). Since an AI software is a combination of software components, AI models, and datasets, a comprehensive AI BOM leverages many other profiles (e.g., Core, Software) to describe the AI System fully (please refer to the examples section).

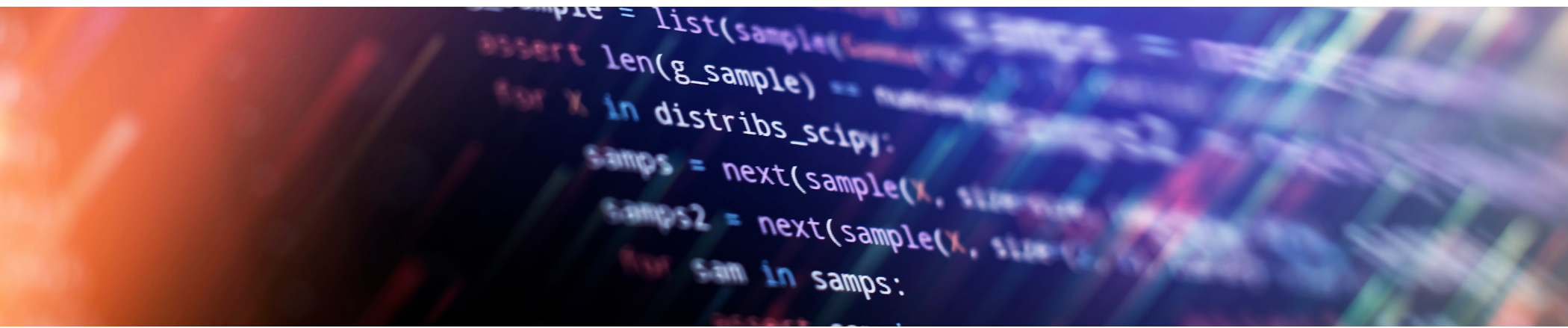

# Developing the SPDX AI and Dataset Profiles

The overarching goal of this initiative is to establish an AI bill of materials (AI BOM) that is not only automated and easy to adopt but also comprehensive enough to describe entire software systems, including AI components. By achieving this, we aim to enable enhanced traceability, license compliance, transparency, and regulatory adherence, thus setting a new standard for AI documentation and management in the tech industry.

Through this structured approach, the working group has laid a robust foundation for a standardized AI and Dataset Profiles, ensuring it meets the current and future demands of AI systems management and governance. Wherever possible, we made an attempt to reuse the fields from other SPDX profiles like Software and Core if fields that one needs to represent an AI software were already available instead of redefining the field. We did so, to avoid redundancy in the SPDX standard and allow for tighter integration with the rest of the standard to enable quicker and wider adoption by the software engineering practitioners. However, these fields, even if borrowed from other profiles, are part of either AI or Dataset Profile.

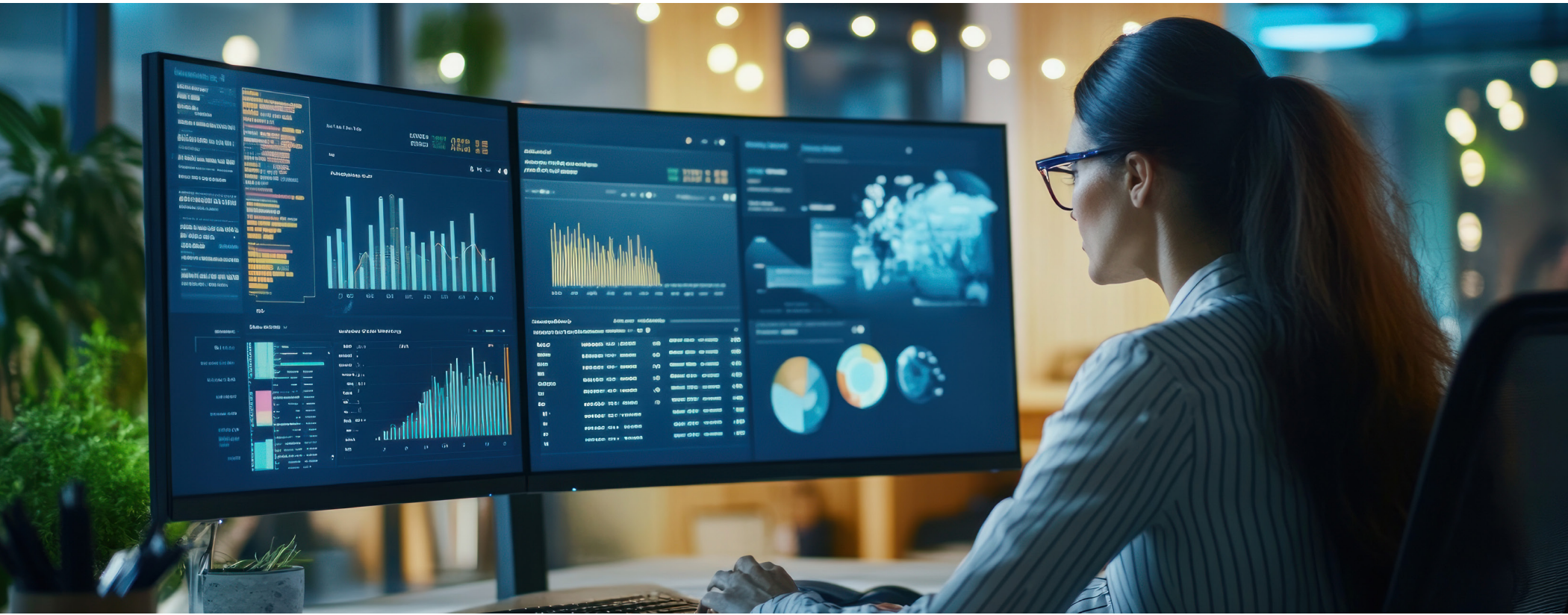

# Defining the AI BOM Framework

## Formation of Working Group and Initial Objectives

In 2021, we established a diverse working group comprising experts from both industry and academia to develop an AI BOM. The group included active researchers from various fields including AI and Software Engineering, professors, chief technology officers, product managers, AI developers, lawyers, and licensing experts, who were primarily recruited through the professional networks of the participants. This open working group convened weekly for one-hour meetings, consistently engaging at least six participants per session. The aim was to craft a standard that not only provided transparency but also enhanced traceability, accountability, provenance, lineage, and automation in AI systems.

## Methodology

Our initial task involved identifying essential fields for the AI BOM. So for the first few months, the working group, drawing from their expertise and diverse background came up with an initial set of fields that would be important to capture in an AI BOM standard. We then analyzed existing tools like model cards, datasheets, and fact sheets with the aim of enhancing the initial list of fields.

The goal of such an analysis was not to ensure that we could capture or represent all the fields that these tools were capturing, rather to ensure that we identify the fields that we have to capture in a AI bill of materials context. Another important consideration that the working group had was that we wanted to ensure the adaptability of our AI BOM standard by practitioners.

Therefore, it was pivotal to balance the urge to capture as many relevant details as possible and ensure that we do not make it too cumbersome or tedious to adopt and use. So there were fields that we decided would be included in future versions and there are only a very small number of fields (explained later in this paper) that were considered mandatory.

Finally, as we mentioned before, we avoided adding fields to AI and Dataset Profiles, if other profiles of SPDX already had the capability to represent any information that is required to arrive at an AI BOM for an AI software. With the aforementioned considerations in context, through meticulous review and democratic discussions within the working group, we analyzed 103 fields proposed in these documents. The inclusion criteria for each field were:

• Relevance to AI BOM requirements

• Existence of a representation method for the field

• Consensus within the group on its inclusion

This iterative process led to the selection of 36 fields, 20 fields for an AI package in the AI Profile (5 of them are required fields) and 18 fields for a dataset package in the Dataset Profile (6 are required), achieved through unanimous consensus, with the report authors serving as tiebreakers when needed.

## Comparison with Datasheets

The following table details the comparison of traditional datasheet fields (we used the Datasheets fields provided in the original paper [r]) with those incorporated in our proposed AI bill of materials, from the AI Profile, the Dataset Profile, and other profiles of SPDX.

| Categories | Sub-categories | In AI BOM? | Matching field (or relationship) in AI BOM or reasons for not including | SPDX profile |
|---|---|---|---|---|
| **Motivation** | For what purpose was the dataset created? | ✔ | intendedUse | Dataset |
| | Who created the dataset and on behalf of which entity? | ✔ | originatedBy | Core |
| | Who funded the creation of the dataset? | ✘ | The comment field can be used in lieu. However, the Working Group decided that this field is not relevant in the context of BOM. | N/A |
| **Composition** | What do the instances that comprise the dataset represent (e.g., documents, photos, people)? | ✔ | datasetType | Dataset |
| | How many instances are there in total (of each type, if appropriate)? | ✔ | datasetSize | Dataset |
| | What data does each instance consist of? | ✔ | datasetType | Dataset |
| | Is any information missing from individual instances? | ✔ | datasetNoise | Dataset |
| | Are there recommended data splits? | ✔ | trainedOn, testedOn relationships | Core |
| | Are there any errors, source of noise, or redundancies in the dataset? | ✔ | datasetNoise | Dataset |

| Categories | Sub-categories | In AI BOM? | Matching field (or relationship) in AI BOM or reasons for not including | SPDX profile |
|---|---|---|---|---|
| Composition | Is the dataset self-contained or does it rely on external resources? | ✔ | downloadLocation, datasetUpdateMechanism | Software, Dataset |
| | Does the dataset contain data that might be considered confidential? | ✔ | confidentialityLevel | Dataset |
| | Does the dataset contain data that might be offensive, insulting, threatening, or anxiety-causing? | ✔ | knownBias | Dataset |
| | Does the dataset identify any sub-populations? | ✔ | knownBias | Dataset |
| | Is it possible to identify individuals directly or indirectly? | ✔ | hasSensitivePersonalInformation | Dataset |
| | Does the dataset have data that is considered sensitive in any way? | ✔ | hasSensitivePersonalInformation | Dataset |
| | Does the dataset contain all possible instances or is it a sample? | ✘ | The Working Group decided that this field is not relevant in the context of BOM. In addition, a dataCollectionProcess field can capture this information. | N/A |
| | Is there a label or target associated with each instance? | ✘ | The Working Group decided that this field is too fine-grained in the context of BOM. | N/A |
| | Are relationships between individual instances made explicit? | ✘ | The Working Group decided that this field is too fine-grained in the context of BOM. | N/A |
| | Does the dataset relate to people? | ✘ | hasSensitivePersonalInformation field can capture aspects of this, but we will refine the field to better capture this information in the future version. | N/A |
| Collection Process | How was the data associated with each instance acquired? | ✔ | dataCollectionProcess | Dataset |

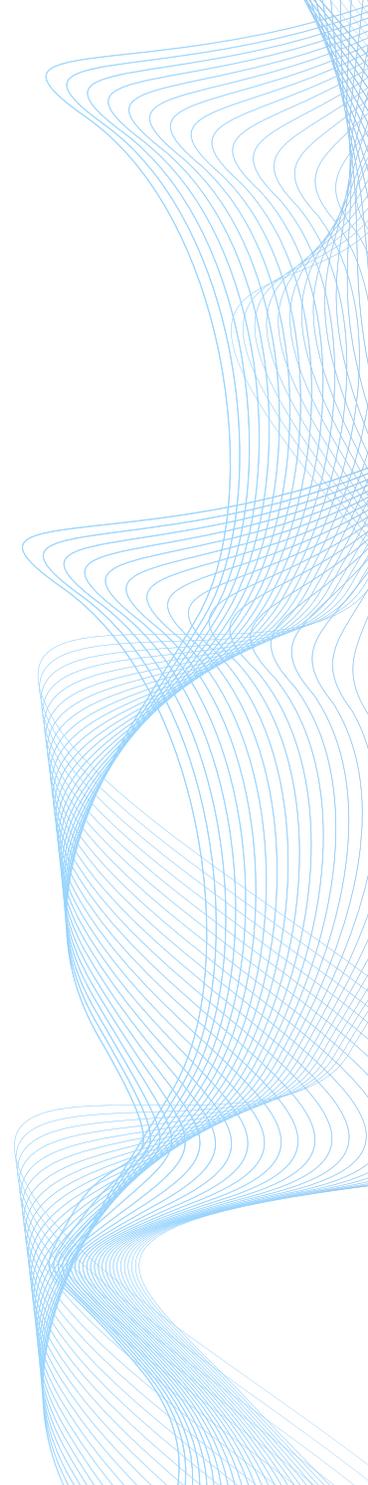

| Categories | Sub-categories | In AI BOM? | Matching field (or relationship) in AI BOM or reasons for not including | SPDX profile |
|---|---|---|---|---|
| Collection Process | What mechanisms or procedures were used to collect the data? | ✔ | dataCollectionProcess | Dataset |
| | Over what timeframe was the data collected? | ✔ | builtTime, releaseTime | Core |
| | Did you collect the data from individuals in question or obtain it via third parties? | ✔ | dataCollectionProcess | Dataset |
| | Were the individuals in question notified about the data collection? | ✔ | dataCollectionProcess | Dataset |
| | Were the individuals in question notified about the data collection? | ✔ | dataCollectionProcess | Dataset |
| | Did the individuals in question consent to the collection and use of their data? | ✔ | dataCollectionProcess | Dataset |
| | If the dataset is a sample from a larger set, what was the sampling strategy? | ✘ | Not included after the Working Group discussion as the dataCollectionProcess field can capture this information. | N/A |
| | Who was involved in the data collection process and how were they compensated? | ✘ | The Working Group decided that this field is not relevant in the context of BOM. | N/A |
| | Were ethical review processes conducted? | ✘ | Not included in this version of the AI BOM. The Working Group decided to consider this for the future versions. | N/A |
| Preprocessing | Was any preprocessing of the data done? | ✔ | dataCollectionProcess | Dataset |
| Uses | Has the dataset been used for any tasks already? | ✔ | intendedUse | Dataset |
| | Is there a repository that links to any or all papers or systems that use the dataset? | ✔ | downloadLocation | Software |

| Categories | Sub-categories | In AI BOM? | Matching field (or relationship) in AI BOM or reasons for not including | SPDX profile |
|---|---|---|---|---|
| Uses | What tasks could the dataset be used for? | ✔ | intendedUse | Dataset |
| | Are there tasks for which the dataset should not be used? | ✘ | Not included in this version of the AI BOM. The Working Group decided to consider this for the future versions. | N/A |
| Distribution | How will the dataset be distributed? | ✔ | downloadLocation | Software |
| | Will the dataset be distributed under a copyright or other IP license or under applicable terms of use? | ✔ | hasDeclaredLicense, hasConcludedLicense relationships | Core |
| | Have any third parties imposed IP-based or other restrictions on the data associated with the instances? | ✔ | hasConcludedLicense relationships | Core |
| | Will the dataset be distributed to third parties outside of the entity on whose behalf it was created? | ✘ | The Working Group decided that this field is not relevant in the context of BOM. | N/A |
| Maintenance | How can the owner/curator/manager of the dataset be contacted? | ✔ | originatedBy | Core |
| | Who will be supporting/hosting/maintaining the dataset? | ✘ | Not included after the Working Group discussion as the supportLevel field can capture this information. | N/A |

**Table 1 - Comparison of SPDX 3.0 with Datasheets**

## Comparison with Model Cards

The following table details the comparison of traditional model cards' fields (we used the Model Cards fields provided in the original paper [p]) with those incorporated in our proposed AI bill of materials, from the AI Profile, the Dataset Profile, and other profiles of SPDX.

| Categories | Sub-categories | In AI BOM? | Matching fields in AI BOM or reasons for not including | SPDX profile associated with the fields |
|---|---|---|---|---|
| **Model Details** | Person or organization developing model | ✔ | suppliedBy | Core |
| | Model date | ✔ | releaseTime | Core |
| | Model version | ✔ | packageVersion | Software |
| | Model type | ✔ | typeOfModel | AI |
| | Information about training algorithms, parameters, fairness constraints or other applied approaches, and features | ✔ | informationAboutTraining, informationAboutApplication | AI |
| | Paper or other resource for more information | ✔ | hasDocumentation | Core |
| | License | ✔ | hasDeclaredLicense, hasConcludedLicense | Core |
| | Where to send questions or comments about the model | ✔ | suppliedBy | Core |

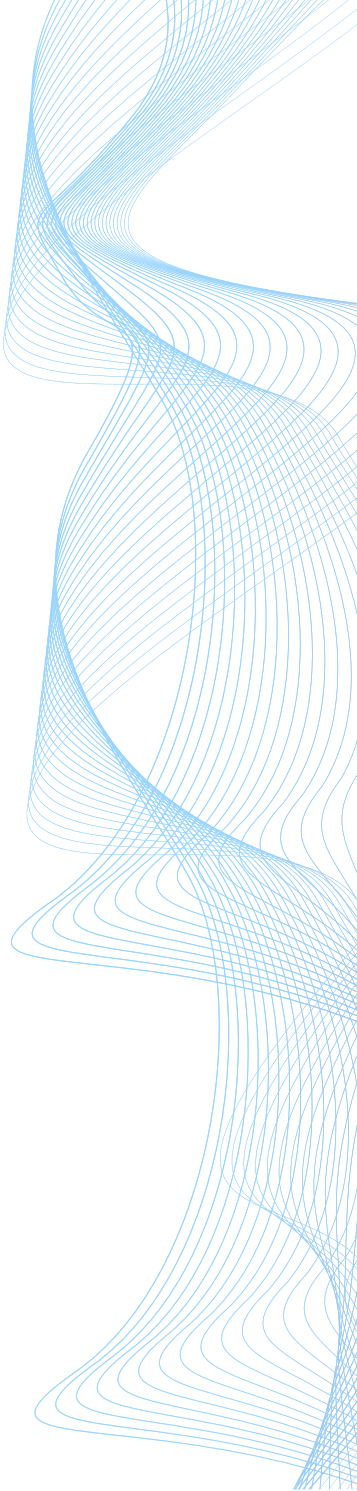

| Categories | Sub-categories | In AI BOM? | Matching fields in AI BOM or reasons for not including | SPDX profile associated with the fields |
|---|---|---|---|---|
| Model Details | Citation details | ✗ | Not included after the Working Group discussion as hasDocumentation relationship can capture this information. Although a more specific relationship can be considered in the future version. | N/A |
| Intended Use | Primary intended uses | ✔ | primaryPurpose | Software |
| | Out-of-scope use cases | ✔ | limitation | AI |
| | Primary intended users | ✗ | The Working Group decided that this field is not relevant in the context of BOM. In addition, intendedUse field can capture this information from a task perspective | N/A |
| Factors | Relevant factors | ✔ | informationAboutTraining, hyperparameter | AI |
| | Evaluation factors | ✔ | informationAboutApplication | AI |
| Metrics | Model performance measures | ✔ | metric | AI |
| | Decision thresholds | ✔ | metricDecisionThreshold | AI |
| | Variation approaches | ✗ | The Working Group decided that this field is not relevant in the context of BOM. | N/A |
| Evaluation Data | Datasets | ✔ | DatasetPackage class; trainedOn, testedOn relationships | Dataset |
| | Motivation | ✔ | intendedUse | Dataset |

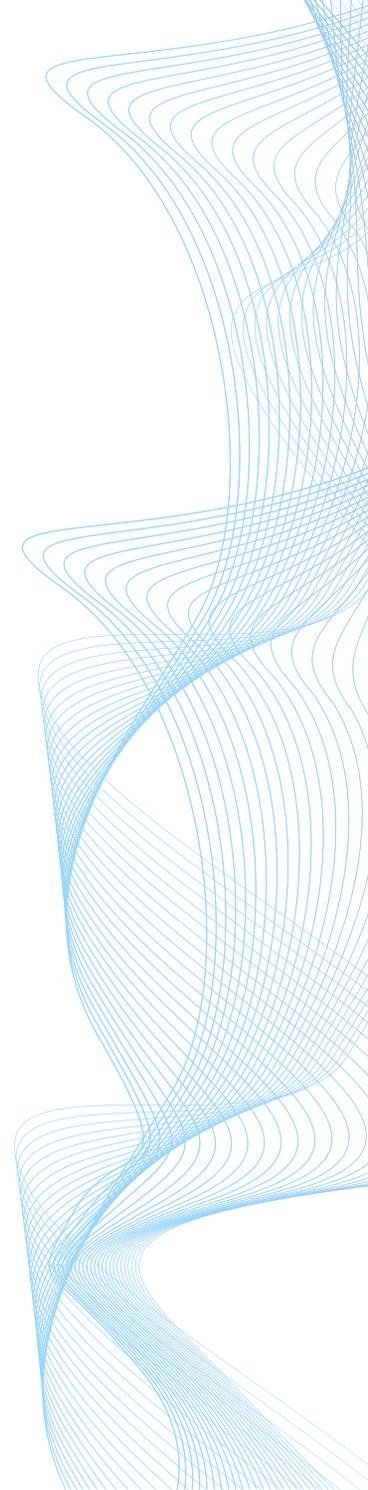

| Categories | Sub-categories | In AI BOM? | Matching fields in AI BOM or reasons for not including | SPDX profile associated with the fields |
|---|---|---|---|---|
| **Evaluation Data** | Preprocessing | ✔ | dataPreprocessing | Dataset |
| **Training Data** | Trained Data | ✔ | trainedOn | Core |
| **Quantitative Analysis** | Unitary results | ✘ | The Working Group decided that this field is not relevant in the context of BOM. | N/A |
| | Intersectional results | ✘ | The Working Group decided that this field is not relevant in the context of BOM. | N/A |
| **Ethical Consideration** | N/A | ✔ | knownBias | Dataset |
| **Recommenda- tions & Caveats** | N/A | ✔ | limitation | AI |

**Table 2 - Comparison of SPDX 3.0 with Model cards**

This table illustrates the thoroughness of our proposed AI BOM in addressing various critical aspects of AI model documenta- tion, pinpointing where it aligns with established practices and where specific enhancements or adjustments have been made.

## Comparison with FactSheets

The following table details the comparison of traditional factsheets' fields (we used the FactSheets fields provided in the original paper [q]) as well as the information about the format that IBM evolved since then [u] with those incorporated in our proposed AI bill of materials, from the AI Profile, Dataset Profile, and other profiles of SPDX.

| Categories | Sub-categories | In AI BOM? | Matching fields in AI BOM or reasons for not including | SPDX profile associated with the fields |
|---|---|---|---|---|
| General | Who are "you" (the supplier) and what type of services do you typically offer (beyond this particular service)? | ✔ | suppliedBy<br>The types of services were not included since the Working Group decided that this field is not relevant in the context of BOM. | Core |
| | What is this service about? | ✔ | informationAboutApplication, primaryPurpose | AI, Software |
| | Describe the outputs of the service | ✔ | informationAboutApplication | AI |
| | What algorithms or techniques does this service implement? | ✔ | typeOfModel;<br>AI profile | AI |
| | Have you updated this FactSheet before? | ✔ | releaseTime | Core |
| | What are the characteristics of the development team? | ✘ | The Working Group decided that this field is not relevant in the context of BOM. | N/A |
| Usage | What is the intended use of the service output? | ✔ | primaryPurpose, domain, informationAboutApplication | AI, Software |
| | What are the key procedures followed while using the service? | ✔ | dataCollectionProcess, dataPreprocessing, informationAboutTraining, informationAboutApplication | AI, Dataset |

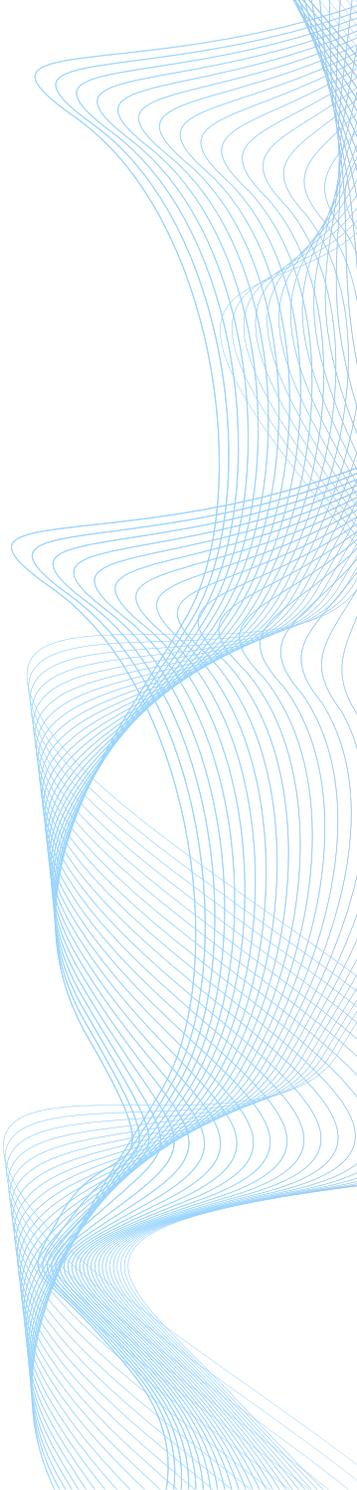

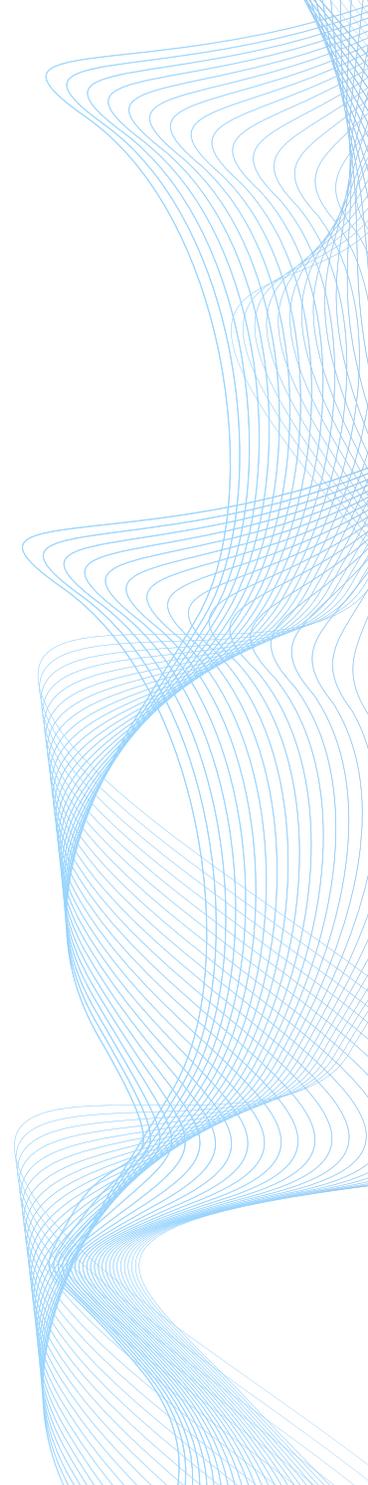

| Categories | Sub-categories | In AI BOM? | Matching fields in AI BOM or reasons for not including | SPDX profile associated with the fields |
|---|---|---|---|---|
| Domains and applications | What are the domains and applications the service was tested on or used for? | ✔ | domain, metric, metricDecisionThreshold; testedOn relationship | AI |
| | How is the service being used by your customers or users? | ✔ | primaryPurpose | Software |
| | Other comments? | ✔ | comment | Core |
| | List applications that the service has been used for in the past. | ✘ | The Working Group decided that this field is not relevant in the context of BOM. | N/A |
| Basic Performance - Testing by service provider | Which datasets was the service tested on? (e.g., links to datasets that were used for testing, along with corresponding datasheets) | ✔ | testedOn relationship | Core |
| | Describe the testing methodology. | ✔ | informationAboutApplication, metric, metricDecisionThreshold; testedOn relationship | AI |
| Basic Performance - Testing by third parties | Describe the test results. | ✔ | metric, metricDecisionThreshold | AI |
| | Is there a way to verify the performance metrics (e.g., via a service API)? | ✔ | metric | AI |
| | Other comments? | ✔ | comment | Core |
| | In addition to the service provider, was this service tested by any third party? | ✘ | Not included in this version of the AI BOM. The Working Group decided to consider this for the future versions. | N/A |
| Safety - General | Are you aware of possible examples of bias, ethical issues, or other safety risks as a result of using the service? | ✔ | knownBias, standardCompliance | AI, Dataset |

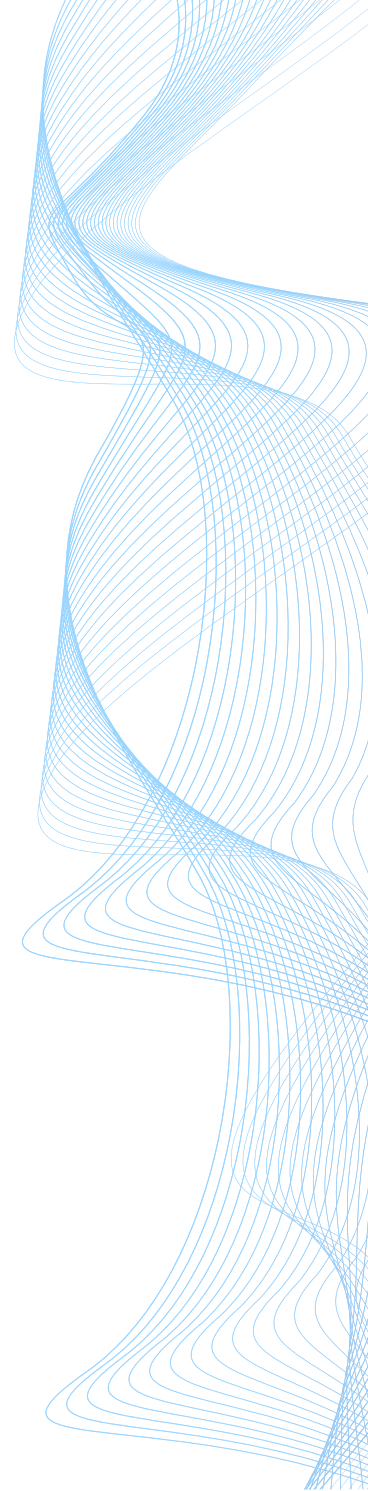

| Categories | Sub-categories | In AI BOM? | Matching fields in AI BOM or reasons for not including | SPDX profile associated with the fields |
|---|---|---|---|---|
| **Safety - General** | Do you use data from or make inferences about individuals or groups of individuals. Have you obtained their consent? | ✔ | knownBias, standardCompliance, confidentialityLevel, limitation | Dataset |
| **Safety - Explainability** | Are the service outputs explainable and/or interpretable? | ✔ | modelExplainability | AI |
| **Safety - Fairness** | For each dataset used by the service: Was the dataset checked for bias? What efforts were made to ensure that it is fair and representative? | ✔ | knownBias, datasetNoise, standardCompliance | AI, Dataset |
| **Safety - Concept Drift** | Does your system make updates to its behavior based on newly ingested data? | ✔ | datasetUpdateMechanism, informationAboutApplication | AI, Dataset |
| | Does the service allow for checking for differences between training and usage data? | ✔ | trainedOn, testedOn relationships; Dataset Profile | Dataset |
| | Other comments? | ✔ | comment | Core |
| | Does the service implement and perform any bias detection and remediation? | ✘ | Not included after Working Group discussion as knownBias field can capture this information and any bias remediation that was conducted can be captured through dataPreprocessing, dataCollectionProcess, modelDataPreprocessing, informationAboutTraining, informationAboutApplication and limitation | N/A |
| | What is the expected performance on unseen data or data with different distributions? | ✘ | Not included after Working Group discussion as domain, intendedUse and primaryPurpose fields can capture this information and the WG deemed that this field was speculative in the context of BOM. | N/A |
| | How is the service tested and monitored for model or performance drift over time? | ✘ | Not included in this version of the AI BOM. The Working Group decided to consider this for the future versions. | N/A |

| Categories | Sub-categories | In AI BOM? | Matching fields in AI BOM or reasons for not including | SPDX profile associated with the fields |
|---|---|---|---|---|
| **Safety - Concept Drift** | How can the service be checked for correct, expected output when new data is added? | ✘ | The Working Group decided that this field is not relevant in the context of BOM. | N/A |
| | Do you test the service periodically? | ✘ | Not included in this version of the AI BOM. The Working Group decided to consider this for the future versions. | N/A |
| **Security** | How could this service be attacked or abused? Please describe. | ✔ | safetyRiskAssessment | AI |
| | List applications or scenarios for which the service is not suitable. | ✔ | intendedUse, primaryPurpose, informationAboutApplication, domain | AI, Dataset |
| | How are you securing user or usage data? | ✔ | anonymizationMethodUsed | Dataset |
| | Other comments? | ✔ | comment | Core |
| | Was the service checked for robustness against adversarial attacks? | ✘ | Not included after the Working Group discussion as safetyRiskAssessment field can capture this information. However, the Working Group decided to consider this for future versions. | N/A |
| | What is the plan to handle any potential security breaches? | ✘ | Not included in this version of the AI BOM. The Working Group decided to consider this for the future versions. | N/A |
| **Lineage - Training Data** | Does the service provide an as-is/canned model? Which datasets was the service trained on? | ✔ | informationAboutTraining, informationAboutApplication, trainedOn | AI |
| | For each dataset: Are the training datasets publicly available? | ✔ | datasetAvailability | Dataset |
| | Did the service require any transformation of the data in addition to those provided in the datasheet? | ✔ | informationAboutTraining, informationAboutApplication, modelDataPreprocessing, dataPreprocessing | AI |

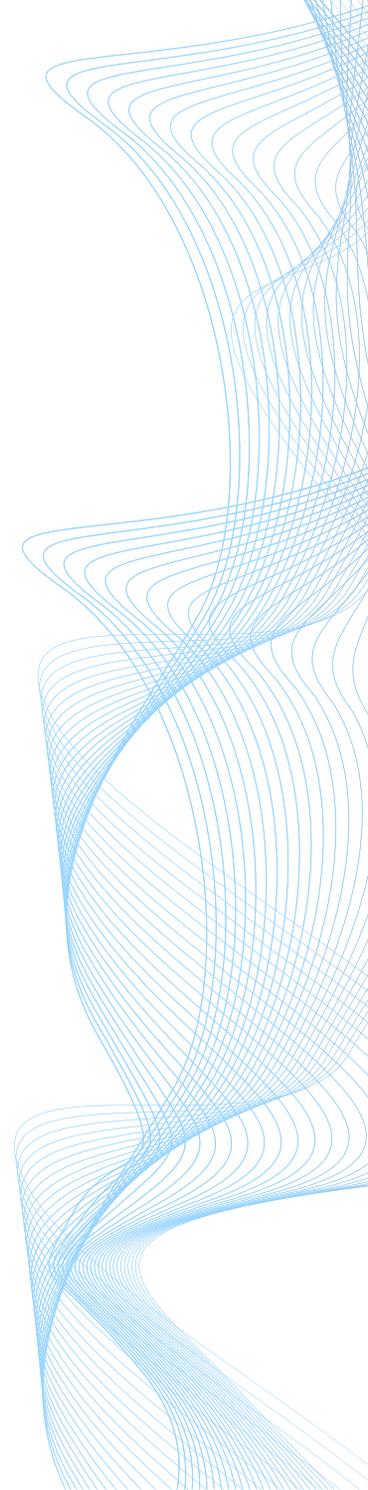

| Categories | Sub-categories | In AI BOM? | Matching fields in AI BOM or reasons for not including | SPDX profile associated with the fields |
|---|---|---|---|---|
| **Lineage - Training Data** | Do you use synthetic data? | ✔ | dataCollectionProcess | Dataset |
| | For each dataset: Does the dataset have a datasheet or data statement? | ✘ | Not included in this version of the AI BOM and is too specific to Factsheets. The Working Group decided to consider this for the future versions. | N/A |
| **Lineage - Trained Models** | How were the models trained? | ✔ | informationAboutTraining | AI |
| | When were the models last updated? | ✔ | releaseTime | Core |
| | Did you use any prior knowledge or re-weight the data in any way before training? | ✔ | informationAboutTraining, modelDataPreprocessing | AI |
| | Other comments? | ✔ | comment | Core |

**Table 3 - Comparison of SPDX 3.0 and IBM FactSheets**

It is important to note that we conducted this analysis in May 2022 and we understand that these tools (i.e., Model cards, FactSheets and Datasheets) have evolved to have more fields (and there are more tools available currently) and as a consequence this analysis may not be the most comprehensive comparison. However, we used this analysis only as an initial starting point and we further evolved the fields captured. The AI and Dataset Profiles of SPDX are being enhanced for the next release to address the evolution in thinking about AI in the community, in line with feedback from various policy makers, practitioners, researchers and developers.

# Using SPDX to comply with International Standards and Regulatory Frameworks

The use case of compliance is shown in this section to ensure that our framework aligns with best practices and current AI standards, We conducted comparisons with ISO and IEEE standards for ethical AI handling. Additionally, our framework was designed to be forward-compatible with the emerging requirements of the EU AI Act, aiming for full compliance with version 3.1. We also validated with both US and EU Medical devices regulations. These validations have helped to align our AI BOM with broader, established guidelines and anticipate future regulatory needs.

The authors of this paper conducted a thorough examination of the European Union Artificial Intelligence Act, the United States Food and Drug Administration medical devices regulations, the IEEE P70xx series of standards for ethical technology, and ISO AI standards. The objective was to assess whether SPDX 3.0 includes fields capable of capturing the details mandated by these regulations and standards. This paper presents the findings of our comparative analysis, evaluating the extent to which SPDX 3.0 AI BOM can represent the required details. This comparison is instrumental in ensuring that our proposed AI BOM aligns with established norms and guidelines in the field. By integrating relevant international standards, we strive to develop a comprehensive and consistent approach to the ethical use of AI and the documentation requirements thereof.

## Ensure Compliance with EU Artificial Intelligence Act[1]

The table presented illustrates the mapping between SPDX 3.0 fields and key clauses of the EU Artificial Intelligence Act (EU AI Act), with a particular focus on the registration requirements for the EU database (Article 49). This registration is mandatory before placing a high-risk AI system on the market or conducting real-world tests. The EU AI Act, effective from 1 August 2024, aims to regulate artificial intelligence by establishing clear rules to ensure that AI systems are designed and deployed in compliance with existing laws and fundamental rights. The legislation categorizes AI systems based on their risk levels and imposes more stringent requirements on high-risk applications to enhance transparency and accountability. By leveraging SPDX 3.0, organizations can effectively map and document their AI systems, ensuring compliance with the EU AI Act's provisions and facilitating a smooth registration process. This integration not only simplifies regulatory adherence but also promotes the development of safe and ethical AI technologies.

---

1   Regulation (EU) 2024/1689 of the European Parliament and of the Council of 13 June 2024 laying down harmonised rules on artificial intelligence and amending Regulations (EC) No 300/2008, (EU) No 167/2013, (EU) No 168/2013, (EU) 2018/858, (EU) 2018/1139 and (EU) 2019/2144 and Directives 2014/90/EU, (EU) 2016/797 and (EU) 2020/1828 (Artificial Intelligence Act) http://data.europa.eu/eli/reg/2024/1689/oj

| Categories | Sub-categories | EU AI Act description and clause | In AI BOM? | Matching field in AI BOM | SPDX profile |
|---|---|---|---|---|---|
| **System, Provider, and Deployer identification** | System unique ID | "A Union-wide unique single identification number of the testing in real world conditions"<br>- Annex IX (1)<br>"URL of the entry of the AI system in the EU database by its provider"<br>- Annex VIII Section C (3)<br>"Any additional unambiguous reference allowing the identification and traceability of the AI system"<br>- Annex VIII Section A (4)<br>- Annex VIII Section B (4 | ✔ | spdxId, externalIdentifier, externalRef, packageUrl in Package | Core |
| | System name | "AI system trade name"<br>- Annex VIII Section A (4)<br>- Annex VIII Section B (4) | ✔ | name in Package (and its subclass AIPackage) | Core |
| | Provider contact | "name, address, and contact details of the provider"<br>- Annex VIII Section A (1)<br>- Annex VIII Section B (1)<br>- Annex VIII Section C (1) | ✔ | createdBy in CreationInfo, together with externalIdentifier in Person and/or Organization | Core |
| | Deployer contact | "name, address, and contact details of the deployer"<br>- Annex VII Section C (1) | ✔ | suppliedBy, together with externalIdentifier in Person and/or Organization | Core |
| **System Details** | Intended purpose | - Annex VIII Section A (5)<br>- Annex VIII Section B (5)<br>- Annex IX (3) | ✔ | primaryPurpose, informationAboutApplication | Core |
| | Information used by system | "Information used by system (data, inputs) and its operating logic"<br>- Annex VIII Section A (6) | ✔ | informationAboutApplication, informationAboutTraining; testedOn, trainedOn relationship; Dataset profile | AI, Dataset |
| | System status | "The status of the AI system (on the market, or in service; no longer placed on the market/in service, recalled)"<br>- Annex VIII Section A (7) | ✔ | validUntilTime, supportLevel | Core |

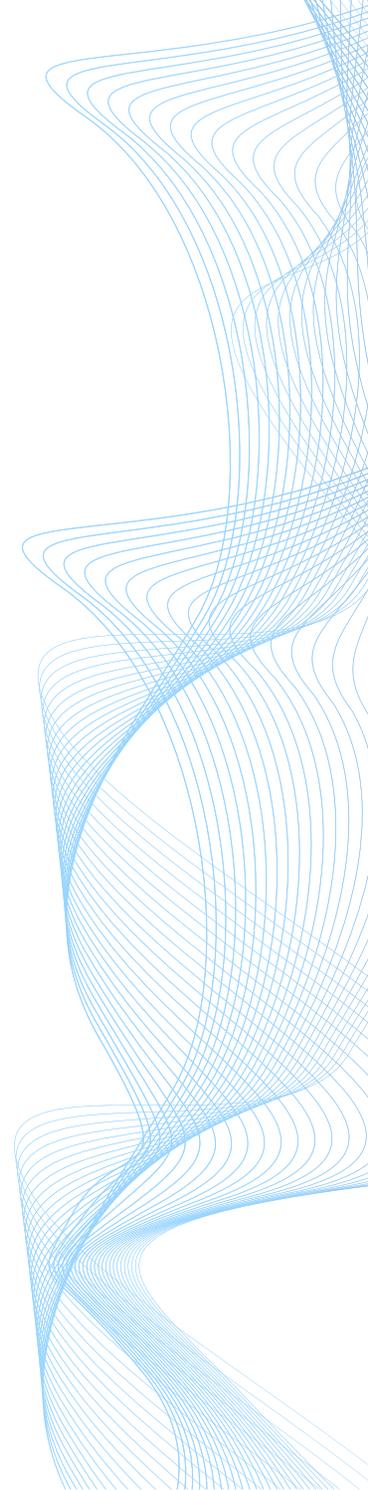

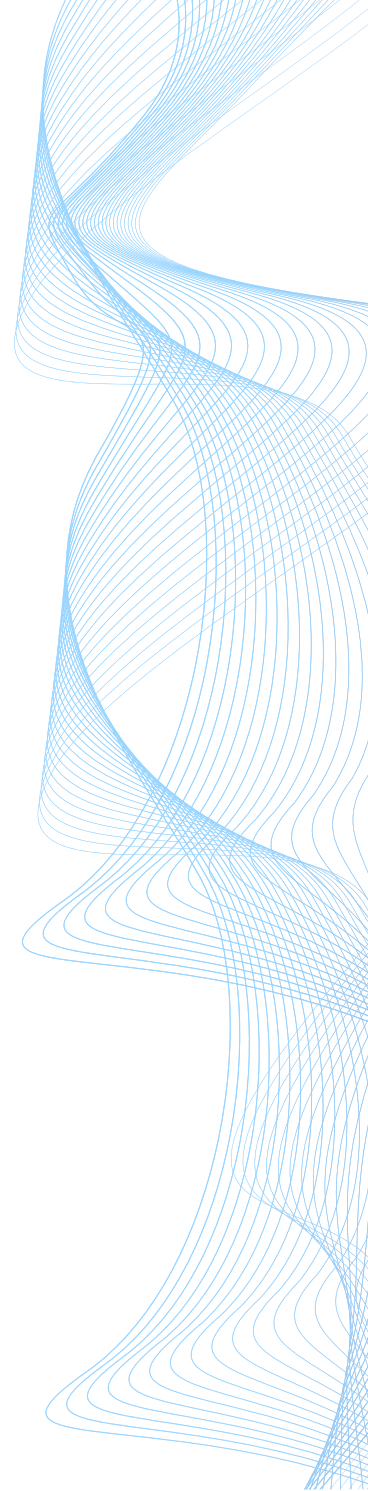

| Categories | Sub-categories | EU AI Act description and clause | In AI BOM? | Matching field in AI BOM | SPDX profile |
|---|---|---|---|---|---|
| **System Details** | System classification | "A short summary of the grounds on which the AI system is considered to be not-high-risk in application of the procedure under Article 6(3)"<br>- Annex VIII Section B (7) | ✔ | informationAboutApplication; hasDocumentation relationship | AI, Core |
| | Instructions for use | "instructions for use for the deployer, and a basic description of the user-interface provided to the deployer, where applicable"<br>- Annex IV (1)(h) | ✔ | informationAboutApplication; hasDocumentation relationship | AI, Core |
| **System Details** | Impact assessment | "A summary of the findings of the fundamental rights impact assessment conducted in accordance with Article 27"<br>- Annex VIII Section C (4)<br>"A summary of the data protection impact assessment carried out in accordance with Article 35 of Regulation (EU) 2016/679 or Article 27 of Directive (EU) 2016/680 as specified in Article 26(8) of this Regulation, where applicable."<br>- Annex VIII Section C (5) | ✔ | hasDocumentation relationship, together with description of the relationship may be used. | Core |
| | Certification | "A copy of the EU declaration of conformity referred to in Article 47"<br>- Annex IV (8)<br>"The type, number and expiry date of the certificate issued by the notified body and the name or identification number of that notified body, where applicable"<br>- Annex VIII Section A (8) | ✔ | standardCompliance, validUntilTime, hasDocumentation relationship, together with description of the relationship | AI, Core |
| **Verification Details** | Testing plan | "A summary of the main characteristics of the plan for testing in real world conditions;"<br>- Annex IX (4) | ✔ | hasDocumentation relationship, together with description of the relationship | Core |

| Categories | Sub-categories | EU AI Act description and clause | In AI BOM? | Matching field in AI BOM | SPDX profile |
|---|---|---|---|---|---|
| Verification Details | Users involved in the testing | "The name and contact details of the provider or prospective provider and of the deployers involved in the testing in real world conditions"<br>- Annex IX (2) | ✘ | There is no specific relationship type for this. However, a Relationship with "to" as Agent (Person and/or Organization) and "relationshipType" as "other" can be used, together with a standardized description value.<br>An externalIdentifier in Person and Organization will hold the contact details. | Core |
| Application Details | Market | "Any Member States in which the AI system has been placed on the market, put into service or made available in the Union"<br>- Annex VIII Section A (10) | ✔ | An Organization can be used, with an externalIdentifier of a standardized country code. | Core |

**Table 4 - EU Artificial Intelligence Act (partial) compliance with SPDX 3.0**

## Ensure Compliance with FDA and EMA Requirements for Medical Devices

The U.S. Food and Drug Administration (FDA) employs software bill of materials (SBOM) to assess the cybersecurity risks of medical devices and evaluate manufacturers' risk management processes. Likewise, the European Medicines Agency (EMA) emphasizes the importance of cybersecurity and risk management for medical devices.

Together with AI and Dataset Profiles, the Source and Build Profiles, also introduced in SPDX 3.0, can effectively identify the requirements for the premarket submission process. Additionally, Runtime Profile is essential for supporting ongoing cybersecurity monitoring and maintenance of medical devices. The Security Profile facilitates the identification and mitigation of newly discovered vulnerabilities, as well as the implementation of software updates and patches.

As of March 22, 2024, relevant FDA guidelines can be accessed at FDA CFR Search. Guidelines for European Union Medical Device Regulation can be accessed at Guidance - MDCG endorsed documents and other guidance.

| Categories | Description and/or CIUAW | In AI BOM? | Matching field in AI BOM | SPDX profile |
|---|---|---|---|---|
| **Package Details** | General understanding of the data and information in the application. [m] | ✔ | informationAboutApplication, Dataset profile | AI, Dataset |
| | Description of condition the device will diagnose, treat, prevent, cure, or mitigate, including a description of the patient population for which the device is intended. | ✔ | primaryPurpose | Software |
| | Description of the foreign and U.S. marketing history, The description shall include the history of the marketing of the device by the applicant and, if known, the history of the marketing of the device by any other person | ✔ | comment | Core |
| | The device, including pictorial representations | ✔ | informationAboutApplication informationAboutTraining | AI |
| **Model Details** | Summary in detail that the reader may gain a general understanding of the data and information in the application. | ✔ | informationAboutApplication informationAboutTraining | AI |
| | Summary of the nonclinical laboratory studies submitted in the application; | ✔ | metric | AI |
| | Include list of all countries in which the device has been marketed and a list of all countries in which the device has been withdrawn from marketing | ✔ | comment | Core |
| **Model Details** | Provide adequate information to demonstrate how the device meets, or justify any deviation from, any performance standard | ✔ | comment | Core |

| Categories | Description and/or ClUAW | In AI BOM? | Matching field in AI BOM | SPDX profile |
|---|---|---|---|---|
| **Data Details** | Description of how the data were collected and analyzed, and a brief description of the results, whether positive, negative, or inconclusive. | ✔ | dataCollectionProcess | Dataset |

**Table 5 - US FDA regulations of medical devices compliance through SPDX 3.0**

| Categories | Description and/or Clause | In SPDX | Field Name | SPDX profile |
|---|---|---|---|---|
| **Package Details** | **Article 10 - General Obligations of Manufacturers:**<br>- Clause 10(1): Manufacturers must ensure that their devices comply with the essential safety and performance requirements set out in Annex I.<br>- Clause 10(2): Manufacturers must establish and maintain a quality management system that covers design, manufacture, and final inspection.<br>- Clause 10(3): Manufacturers must draw up and keep up to date the technical documentation for their devices. | ✔ | spdxId, locator, created, packageVersion, createdBy, originatedBy, suppliedBy, downloadLocation, primaryPurpose, datasetType, verifiedUsing, typeOfModel, informationAboutApplication, informationAboutTraining, metric, comment, description<br><br>hasConcludedLicense, hasDeclaredLicense, testedOn, trainedOn relationships | AI, Core, Dataset, Licensing, Software |

**Table 6 - EU Medical Device Regulation (MDR) compliance through SPDX 3.0**

## Ensure compliance with IEEE Ethical Technology Standards (P70xx series)

The IEEE P70xx series standards delineate the essential information that must be documented to ensure compliance. Although not all fields required by the P70xx standards are explicitly mandatory for the SPDX, adherence to the P70xx standard necessitates the inclusion of all required fields from SPDX 3.0. It is imperative to comprehensively document all relevant information to maintain compliance with both standards and ensure the integrity and reliability of AI systems. SPDX 3.0 offers a comprehensive framework for documenting all necessary information, making it an ideal choice for organizations aiming to meet both IEEE P70xx and SPDX standards.

| Categories | Description and/or Section | Field in SPDX | Field Name | SPDX Profile |
|---|---|---|---|---|
| Package Details | 7000 - Clause 1.2: mandates that these ethical values and principles must be considered throughout the system design process.<br>7001 - Clause 2.2: mandates that stakeholders must be involved in the ethical considerations and decision-making processes.<br>7002 - Clause 5.2: mandates that the audit trails must include information about who created the data, when it was created, and any subsequent modifications.<br>7009 - Clause 1.1: requires identification documentation of system architecture principles and ownership that support fail-safe design.<br>7014 - Clause 5.1: requires the establishment of accountability mechanisms to ensure that those responsible for the system are held accountable for its ethical design and operation | ✔ | createdBy | Core |
| | 7000 - Clause 3.2: mandates that the ethical impact assessment must be documented<br>7001 - Clause 4.2: mandates that the ethical impact assessment must be documented<br>7002 - Clause 2.2: mandates that the privacy impact assessment must document the sources of data, including information about who created the data. | ✔ | impactStatement | Security |

| Categories | Description and/or Section | Field in SPDX | Field Name | SPDX Profile |
|---|---|---|---|---|
| Package Details | 7009 - Clause 5.2: mandates the documentation of ethical considerations and their impact on system design and operation<br><br>7010 - Clause 8.1: requires the reporting of wellbeing metrics and their impact on users and society<br><br>7014 - Clause 3.2: mandates that the ethical impact assessment must be documented and reviewed by stakeholders, including information about who created the system components | ✔ | impactStatement | Security |
| | 7000 - Clause 2.2: mandates that these transparency and explainability requirements must be documented and communicated to stakeholders, including information about who supplied the system components.<br><br>7001 - Clause 1.2: mandates that these ethical principles and values must be considered throughout the system life cycle, including the documentation of who supplied the system components<br><br>7002 - Clause 4.2: mandates that the documentation of data provenance must include details about who created the data, when it was created, and any subsequent modifications<br><br>7005 - Clause 5.2: mandates that the audit trails must include information about who supplied the data, when it was supplied, and any subsequent modifications<br><br>7007 - Clause 5.1: requires the establishment of accountability mechanisms to ensure that those responsible for the system are held accountable for its ethical design and operation<br><br>7009 - Clause 5.2: mandates that these accountability mechanisms must be documented and reviewed periodically, including information about who supplied the system components.<br><br>7010 - Clause 5.1: requires the establishment of accountability mechanisms to ensure that those responsible for the system are held accountable for its ethical design and operation<br><br>7014 - Clause 5.2: mandates that these accountability mechanisms must be documented and reviewed periodically, including information about who supplied the system components | ✔ | suppliedBy | Core |

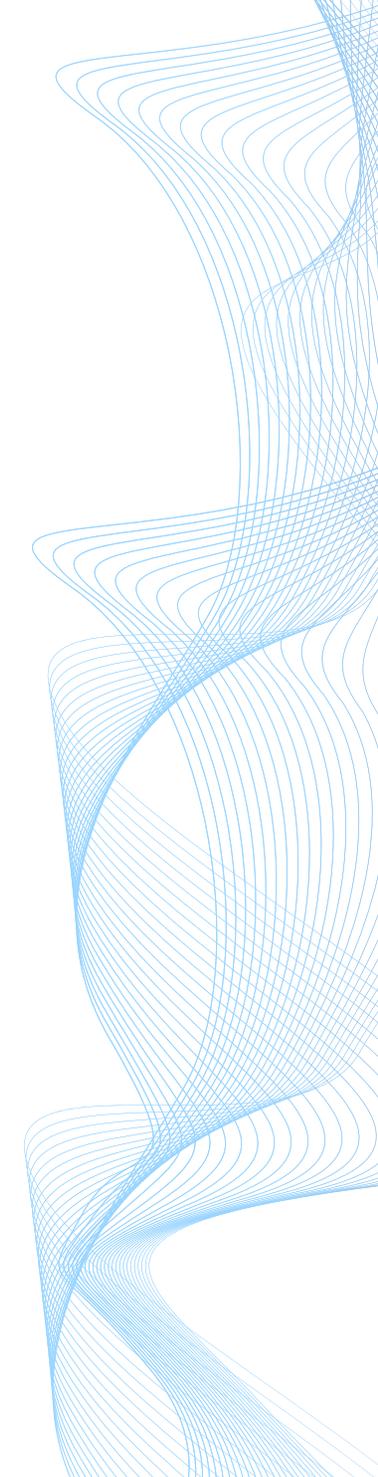

| Categories | Description and/or Section | Field in SPDX | Field Name | SPDX Profile |
|---|---|---|---|---|
| Package Details | 7014 - Clause 11.1: requires the establishment of processes for ongoing monitoring and maintenance of the system to address any emerging ethical issues. This could include monitoring data lifecycle management practices and ensuring that data expiry policies are followed. | ✔ | validUntilTime | Core |
| | 7000 - Clause 5.2: mandates that these strategies must be documented and reviewed to ensure their effectiveness<br><br>7001 - Clause 3.2: mandates that these transparency requirements must be documented<br><br>7002 - Clause 3.2: mandates that these mechanisms must include provisions for documenting the purpose and context of data usage, which may include a description of the application or system<br><br>7005 - Clause 2.1: requires the establishment of transparency and accountability mechanisms to ensure that employer data is used ethically and responsibly<br><br>7009 - Clause 2.2: mandates the development and implementation of recovery mechanisms that can restore system functionality or transition to a safe state in the event of a failure<br><br>7010 -- Clause 7.2: mandates the establishment of monitoring and evaluation processes to assess the effectiveness of the wellbeing metrics<br><br>7014 - Clause 6.2: mandates that this documentation must be made available to stakeholders and regulatory authorities as needed, including a description of the application or system. | ✔ | description | Core |
| | 7000 - Clause 5.2: mandates that these strategies must be documented and reviewed to ensure their effectiveness<br><br>7001 - Clause 5.2: mandates that these strategies must be documented and reviewed to ensure their effectiveness<br><br>7002 - Clause 1.2: mandates that these policies and procedures must include provisions for documenting the purpose and context of data usage, which may include the primary purpose of the AI application. | ✔ | primaryPurpose | Software |

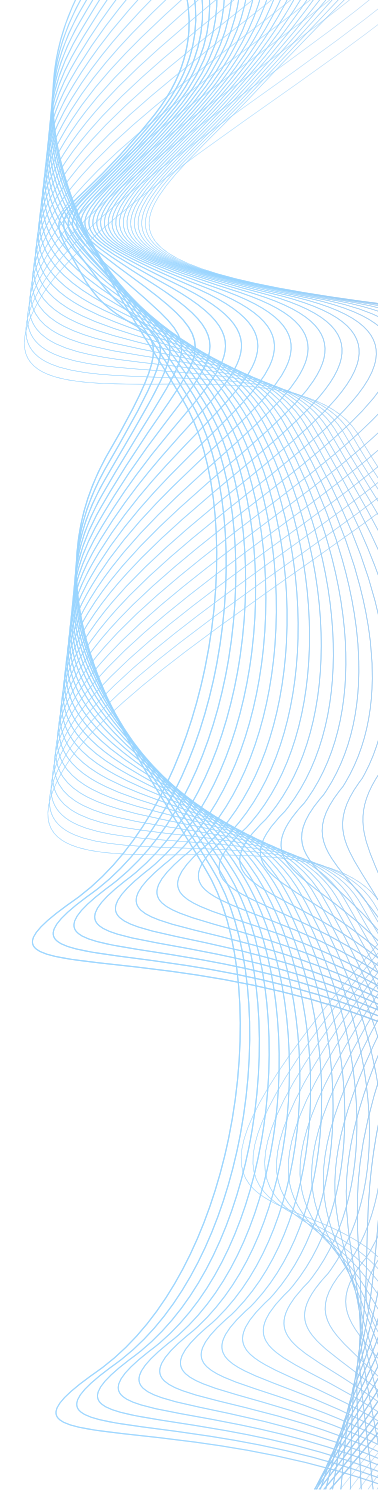

| Categories | Description and/or Section | Field in SPDX | Field Name | SPDX Profile |
|---|---|---|---|---|
| Package Details | 7010 - Clause 1.2: outlines the objectives of the metrics, including the promotion of human wellbeing, ethical considerations, and the responsible use of AI.<br><br>7014 - Clause 6.2: mandates that this documentation must be made available to stakeholders and regulatory authorities as needed, including a description of the primary purpose of the system | ✔ | primaryPurpose | Software |
|  | 7000 - Clause 5.2: mandates that these accountability mechanisms must be documented and reviewed periodically, including information about where data or system components can be downloaded<br><br>7001 - Clause 2.2: mandates that these transparency and explainability requirements must be documented and communicated to stakeholders, including information about where data or system components can be downloaded.<br><br>7002 - Clause 3.2: mandates that these mechanisms must include provisions for documenting the provenance of data, including information about where data can be downloaded.<br><br>7005 - Clause 5.2: mandates that the audit trails must include information about who supplied the data, when it was supplied, and any subsequent modifications.<br><br>7007 - Clause 6.2: mandates that this documentation must be made available to stakeholders and regulatory authorities as needed, including information about where data or system components can be downloaded.<br><br>7014 - Clause 5.2: mandates that these accountability mechanisms must be documented and reviewed periodically, including information about where data or system components can be downloaded. | ✔ | downloadLocation | Software |
| AI Details | 7014 - Clause 5.2: The standard mandates that these accountability mechanisms must be documented and reviewed periodically, including information about the type of model used in the system. | ✔ | typeOfModel | AI |

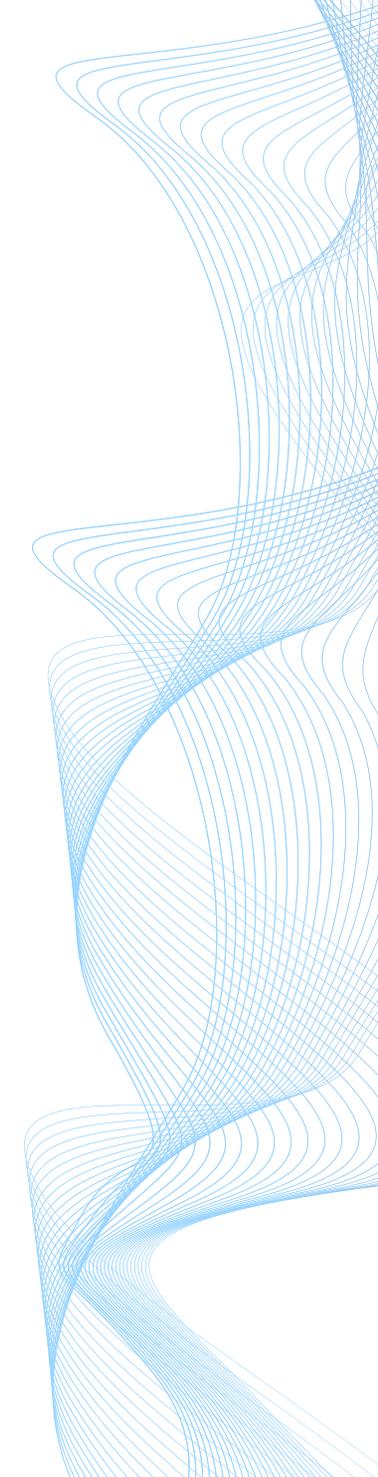

| Categories | Description and/or Section | Field in SPDX | Field Name | SPDX Profile |
|---|---|---|---|---|
| AI Details | 7000 - Clause 5.2: mandates that these strategies must be documented and reviewed to ensure their effectiveness.<br><br>7002 - Clause 2.1: requires the conduct of privacy impact assessments to identify potential privacy risks associated with data processing activities.<br><br>7010 - Clause 8.2: mandates that the reporting must be transparent, accessible, and understandable to stakeholders<br><br>7014 - Clause 4.2: mandates that these transparency and explainability requirements must be documented and communicated to stakeholders. | ✔ | modelExplainability | AI |
| | 7005 - Clause 5.2: mandates that the audit trails must include information about who supplied the data, when it was supplied, and any subsequent modifications<br><br>7010 - Clause 7.3: requires the documentation of implementation and monitoring practices<br><br>7014 - Clause 6.2: mandates that this documentation must be made available to stakeholders and regulatory authorities as needed, including comprehensive information about the AI application. | ✔ | informationAboutApplication | AI |
| | 7005 - Clause 5.2: mandates that the audit trails must include information about who supplied the data, when it was supplied, and any subsequent modification<br><br>7009 - Clause 6.1: requires the implementation of data management practices that ensure the integrity and security of data used by the system<br><br>7002 - Clause 1.2: mandates that these policies and procedures must include provisions for documenting the provenance of data, including information about who created the data.<br><br>7014 - Clause 4.2: mandates that these policies and procedures must be documented and communicated to stakeholders, including comprehensive information about the AI application training | ✔ | informationAboutTraining | AI |

| Categories | Description and/or Section | Field in SPDX | Field Name | SPDX Profile |
|---|---|---|---|---|
| AI Details | 7001 - Clause 8.2: mandates that the verification and validation processes must be documented and reviewed by stakeholders<br><br>7009 - Clause 2.3: requires the documentation of fault detection and recovery mechanisms, including their effectiveness and limitations.<br><br>7014 - Clause 5.2: mandates that these strategies must be documented and reviewed to ensure their effectiveness. | ✔ | limitation | AI |
| | 7000 - Clause 6.2: mandates that the verification and validation processes must be documented<br><br>7001 - Clause 8.2:mandates that the verification and validation processes must be documented<br><br>7009 - Clause 3.2: mandates the validation and verification of the system to ensure that it meets safety requirements<br><br>7010 - Clause 2.2: document the various types of wellbeing metrics, including physical, mental, social, and environmental wellbeing<br><br>7014 - Clause 10.1: requires the verification and validation of the system to ensure that it meets the defined ethical and transparency requirements | ✔ | metric | AI |
| | 7014 - Clause 10.2: verification and validation of the system to ensure that meets expected quality metrics | ✔ | metricDecisionThreshold | AI |
| | 7000 - Clause 7.2: mandates that support processes must be documented and reviewed periodically.<br><br>7001 - Clause 9.2: mandates that these processes must be documented and reviewed periodically.<br><br>7009 - Clause 3.3: The standard requires the establishment of ongoing monitoring and maintenance processes to address emerging risks and vulnerabilities. | ✔ | supportLevel | Core |
| | 7000 - Clause 10.2: mandates that compliance with laws and regulations must be documented and reviewed<br><br>7001 - Clause 11.2: mandates that compliance with laws and regulations must be documented and reviewed. | ✔ | standardCompliance | AI |

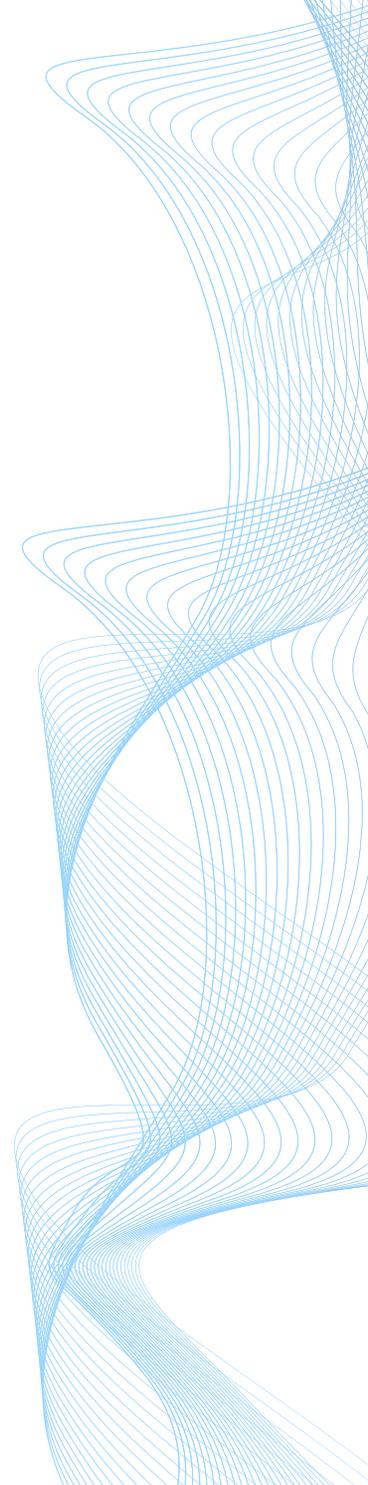

| Categories | Description and/or Section | Field in SPDX | Field Name | SPDX Profile |
|---|---|---|---|---|
| **AI Details** | 7002 - Clause 2.2: mandates that these transparency and explainability requirements must be documented and communicated to stakeholders, including information about compliance with relevant laws and regulations<br><br>7005 - Clause 6.2: mandates that this documentation must be made available to stakeholders and regulatory authorities as needed, including information about compliance with relevant laws and regulations.<br><br>7007 - Clause 4.2: mandates that these policies and procedures must be documented and communicated to stakeholders, including information about compliance with relevant laws and regulations.<br><br>7009 - Clause 8.2: mandates the documentation of compliance with laws and regulations<br><br>7010 - Clause 5.2: mandates that the data used to measure wellbeing metrics must be managed securely and in compliance with relevant regulations | ✔ | standardCompliance | AI |
| | 7000 - Clause 4.2: mandates that these ethical requirements must be integrated into the system design and development process.<br><br>7009 - Clause 3.1: Requires the conduct of risk assessments to identify potential failure modes and their impacts.<br><br>7014 - Clause 4.1: requires the identification of potential risks associated with the design and operation of systems emulating empathy. | ✔ | safetyRiskAssessment | AI |
| **Software Details** | 7000, 7001, 7002, 7005, 7007<br><br>7010, 7014 - requires the i domain type ie. context of the specific application domain. | ✔ | domain | AI |
| | 7000, 7001, 7002, 7005, 7007<br><br>7010, 7014 - Clause 6.1: requires the establishment of mechanisms for obtaining informed consent from users regarding the use of emulated empathy. | ✔ | comment | Core |

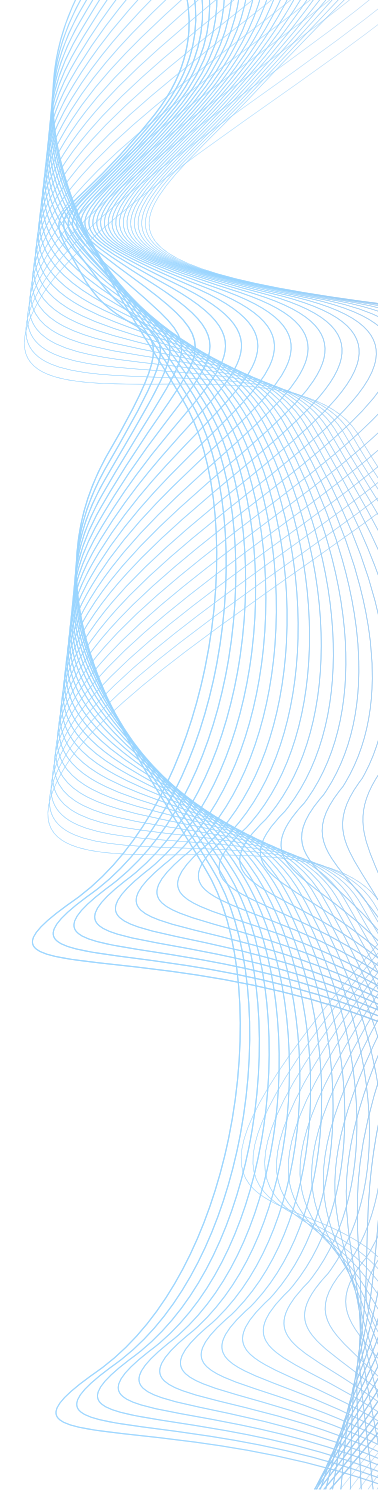

| Categories | Description and/or Section | Field in SPDX | Field Name | SPDX Profile |
|---|---|---|---|---|
| **Software Details** | 7005 - Clause 3.2: mandates that the documentation of data provenance must include details about who supplied the data, when it was supplied, and any subsequent modifications.<br><br>7009 - Clause 6.1: The standard requires the implementation of data management practices that ensure the integrity and security of data used by the system.<br><br>7010 - Clause 5.1: requires the establishment of data collection methods that are ethical, transparent.<br><br>7014 - Clause 8.2: mandates all data policies and procedures must be documented. | ✔ | modelDataPreprocessing | AI |
| | 7014 - Clause 5.2: mandates that sensitive data must be documented. | ✔ | useSensitivePersonalInformation | AI |
| | 7005 - Clause 3.2: mandates that the documentation of data provenance must include details about who supplied the data, when it was supplied, and any subsequent modifications.<br><br>7009 - Clause 6.1: requires the implementation of data management practices that ensure the integrity and security of data.<br><br>7010 - Clause 5.3: requires the documentation of data collection and management practices.<br><br>7014 - Clause 6.2: mandates that these mechanisms must respect user autonomy and provide options for users to control the level of emulated empathy. | ✔ | autonomyType | AI |
| | 7009 - Clause 7.1: requires the assessment of the environmental impact of the system, including energy consumption and waste management. & Clause 7.2: mandates the design of the system to operate safely under various environmental conditions.<br><br>7014 - Clause 4.2: mandates that these policies and procedures must be documented and communicated to stakeholders, which could include considerations about energy use if relevant. | ✔ | energyConsumption | AI |
| **Data Details** | 7005 - Clause 3.2: mandates that the documentation of data provenance must include details about who supplied the data, when it was supplied, and any subsequent modifications | ✔ | anonymizationMethodUsed | Dataset |

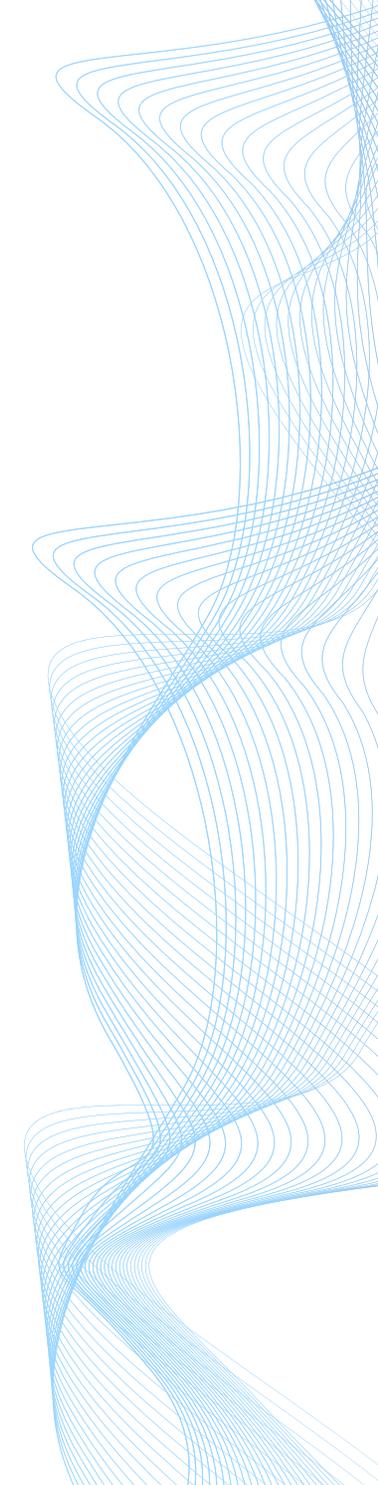

| Categories | Description and/or Section | Field in SPDX | Field Name | SPDX Profile |
|---|---|---|---|---|
| **Data Details** | 7009 - Clause 6.1: requires the implementation of data management practices that ensure the integrity and security of data used by the system.<br><br>7014 - Clause 8.2: mandates all data policies and procedures must be documented | ✔ | anonymizationMethodUsed | Dataset |
| | 7001 - Clause 6.2: mandates that data governance policies and procedures must be documented and communicated to stakeholders.<br><br>7002 - Clause 1.2: mandates that these policies and procedures must include provisions for documenting the provenance of data, including information about who created the data.<br><br>7005 - Clause 3.2: mandates that the documentation of data provenance must include details about who supplied the data, when it was supplied, and any subsequent modifications<br><br>7009 - Clause 6.1: requires the implementation of data management practices that ensure the integrity and security of data used by the system.<br><br>7010 - Clause 5.3: requires the documentation of data collection and management practices.<br><br>7014 - Clause 8.2: mandates all data policies and procedures must be documented | ✔ | dataCollectionProcess | Dataset |
| | 7002 - Clause 1.2: mandates that these policies and procedures must include provisions for documenting the provenance of data, including information about who created the data.<br><br>7005 - Clause 3.2: mandates that the documentation of data provenance must include details about who supplied the data, when it was supplied, and any subsequent modifications<br><br>7009 - Clause 6.1: requires the implementation of data management practices that ensure the integrity and security of data used by the system.<br><br>7010 - Clause 5.3: requires the documentation of data collection and management practices.<br><br>7014 - Clause 8.2: mandates all data policies and procedures must be documented | ✔ | dataPreprocessing | Dataset |

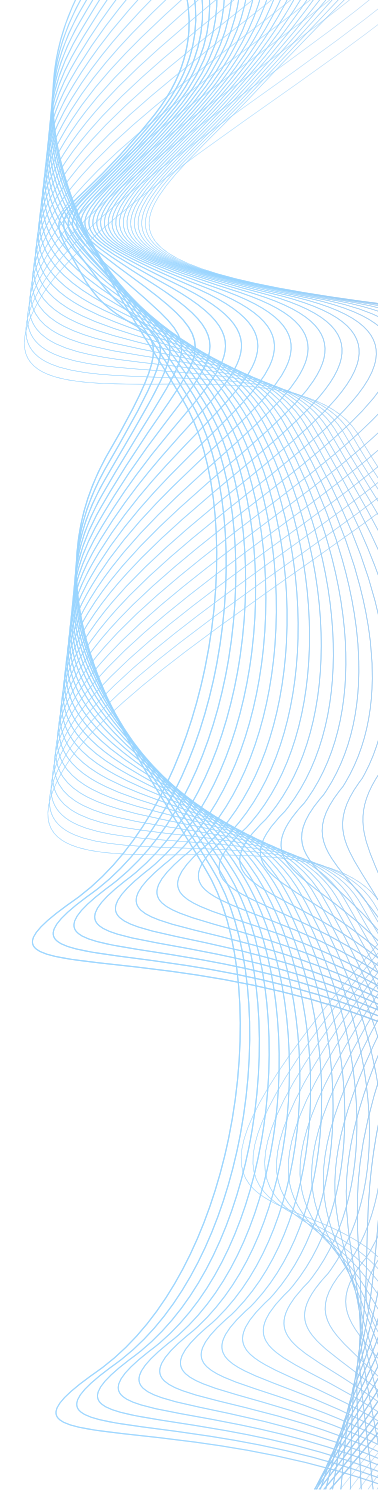

| Categories | Description and/or Section | Field in SPDX | Field Name | SPDX Profile |
|---|---|---|---|---|
| **Data Details** | 7002 - Clause 1.2: mandates that these policies and procedures must include provisions for documenting the provenance of data, including information about who created the data<br><br>7009 - Clause 6.1: requires the implementation of data management practices that ensure the integrity and security of data used by the system.<br><br>7014 - Clause 8.2: mandates all data policies and procedures must be documented | ✔ | datasetAvailability | Dataset |
| | 7002 - Clause 1.2: mandates that these policies and procedures must include provisions for documenting the provenance of data, including information about who created the data<br><br>7005 - Clause 3.2: mandates that the documentation of data provenance must include details about who supplied the data, when it was supplied, and any subsequent modifications<br><br>7009 - Clause 6.1: requires the implementation of data management practices that ensure the integrity and security of data used by the system.<br><br>7010 - Clause 5.3: requires the documentation of data collection and management practices.<br><br>7014 - Clause 8.2: mandates all data policies and procedures must be documented | ✔ | datasetNoise | Dataset |
| | 7002 - Clause 1.2: mandates that these policies and procedures must include provisions for documenting the provenance of data, including information about who created the data<br><br>7005 - Clause 3.2: mandates that the documentation of data provenance must include details about who supplied the data, when it was supplied, and any subsequent modifications<br><br>7009 - Clause 6.1: requires the implementation of data management practices that ensure the integrity and security of data used by the system.<br><br>7014 - Clause 8.2: mandates all data policies and procedures must be documented | ✔ | datasetSize | Dataset |

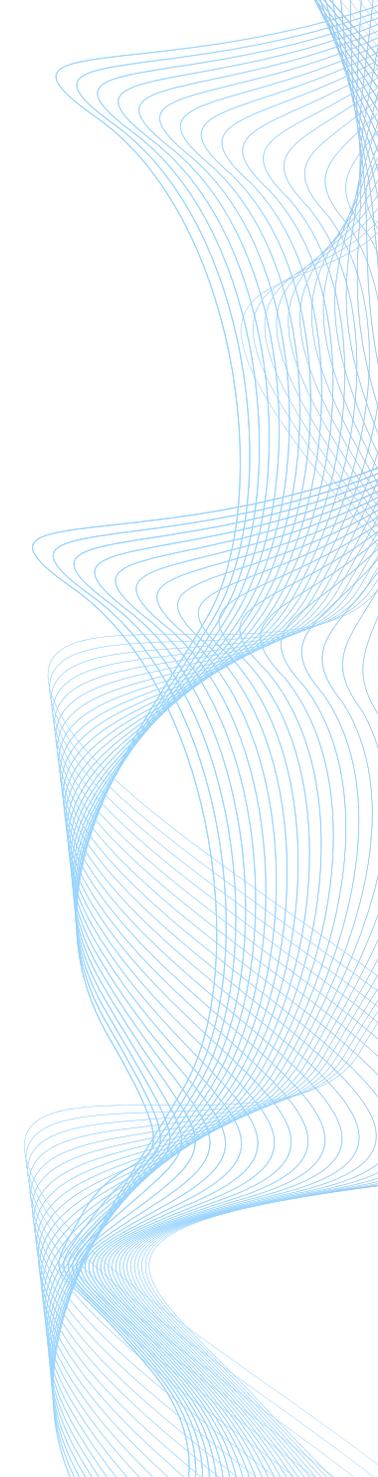

| Categories | Description and/or Section | Field in SPDX | Field Name | SPDX Profile |
|---|---|---|---|---|
| **Data Details** | 7002 - Clause 1.2: mandates that these policies and procedures must include provisions for documenting the provenance of data, including information about who created the data<br><br>7005 - Clause 3.2: mandates that the documentation of data provenance must include details about who supplied the data, when it was supplied, and any subsequent modifications<br><br>7009 - Clause 6.1: requires the implementation of data management practices that ensure the integrity and security of data used by the system.<br><br>7014 - Clause 8.2: mandates all data policies and procedures must be documented | ✔ | datasetType | Dataset |
| | 7009 - Clause 6.1: requires the implementation of data management practices that ensure the integrity and security of data used by the system.<br><br>7010 - Clause 5.3: requires the documentation of data collection and management practices.<br><br>7014 - Clause 8.2: mandates all data policies and procedures must be documented | ✔ | datasetUpdateMechanism | Dataset |
| | 7000 - Clause 6.2: mandates intended use must be documented<br><br>7001 - Clause 6.2: mandates that data governance policies and procedures must be documented<br><br>7002 - Clause 5.2: mandates that these practices must include provisions for documenting the intended use of data and obtaining explicit consent for such use.<br><br>7005 - Clause 5.2: mandates that the audit trails must include information about the intended use of employer data.<br><br>7007 - Clause 6.2: The standard mandates that this documentation must be made available to stakeholders and regulatory authorities as needed, including information about the intended use of the system<br><br>7009 - Clause 6.2: mandates the protection of the system from cyber threats and the establishment of cybersecurity measures<br><br>7010 - Clause 1.2: mandates that these ethical principles and values must be considered throughout the system life cycle, including the documentation of the intended use of the system.<br><br>7014 - Clause 8.1: requires the establishment of data governance policies and procedures to ensure the ethical and responsible use of data in emulated empathy systems. | ✔ | intendedUse | Dataset |

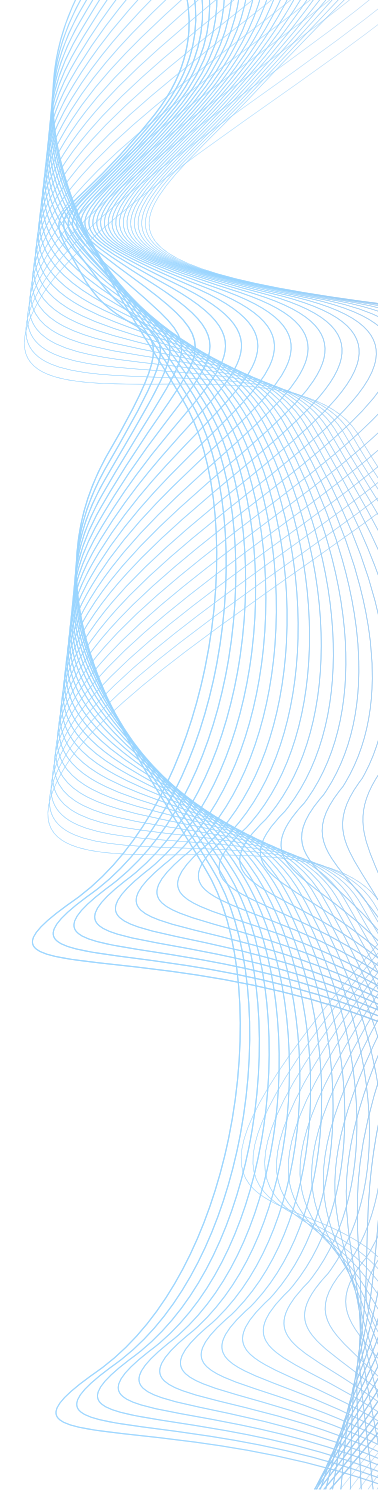

| Categories | Description and/or Section | Field in SPDX | Field Name | SPDX Profile |
|---|---|---|---|---|
| **Data Details** | 7001- Clause 6.2: mandates that data governance policies and procedures must be documented<br><br>7002 - Clause 1.2: mandates that these policies and procedures must include provisions for documenting the provenance of data, including information about who created the data<br><br>7009- Clause 6.1: requires the implementation of data management practices that ensure the integrity and security of data used by the system.<br><br>7010 - Clause 5.3: requires the documentation of data collection and management practices.<br><br>7014 - Clause 7.2: mandates that fairness principles must be integrated into the system design and operation. | ✔ | knownBias | Dataset |
| | 7014 - Clause 5.2: mandates that empathetic/sensitive data must be documented | ✔ | hasSensitivePersonalInformation | Dataset |
| | 7014 - Clause 8.2: mandates all data policies and procedures must be documented | ✔ | sensor | Dataset |

**Table 7 - Comparison of IEEE P70xx publicy available standards and SPDX 3.0**

This table compares multiple IEEE P70xx ethical technology standards that are publicly available to the fields that SPDX must capture to be compliant. Furthermore, we are actively working with various regulatory bodies, including the US Executive Order 14028 consortium and the EU AI Office, as well as AI and data committees in Canada, Japan, and the UK, and multiple international standard organizations (IEEE and ISO) to validate SPDX 3.0 coverage.

# Ongoing Revisions and Updates

As the AI field continues to advance, our working group remains diligent in identifying and integrating new fields and considerations into the AI bill of materials (AI BOM). Regular updates and revisions are integral to our process, ensuring that the AI BOM remains current and adheres to the highest standards of ethical AI use and documentation. For example, we have taken into account the emerging needs outlined in the EU AI Act.

Similarly, we are aware that the latest version of model cards on Hugging Face [s] includes fields that we did not consider such as "fine-tuned from model," "funded by," "language," "demo," "bias recommendations," "testing factors," "model examination," and a richer set of fields for capturing environmental impact and compute infrastructure. We aim to incorporate these fields into our standard with the next revision after thorough discussion with the community regarding their relevance, usefulness, and practicality, as the objectives of these documentation standards differ from ours.

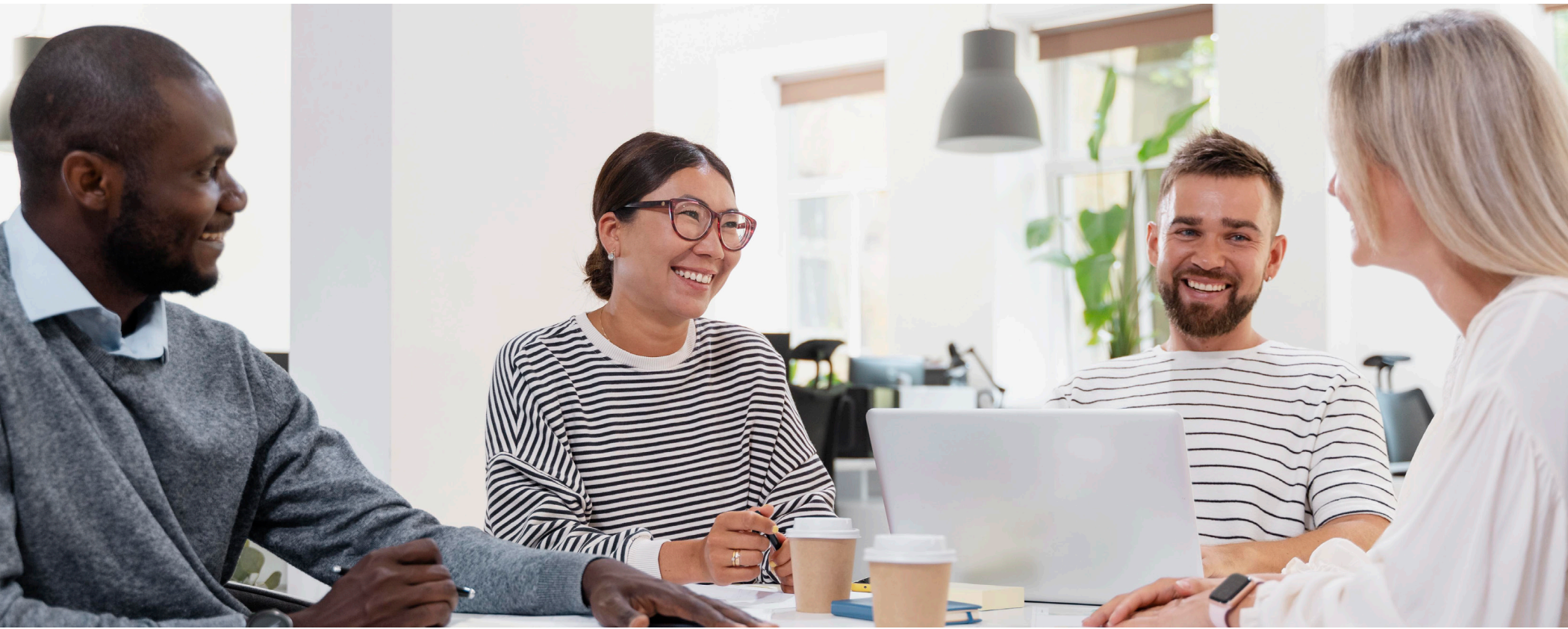

# Key Elements for the Profiles

**What are the key elements in the SPDX AI and Dataset Profiles**

**Licensing details:** SPDX licensing provides a standardized way of communicating licensing information about software and other content, and in our case including AI model and datasets' license information, making it easier for organizations to manage their open-source software supply chain and comply with license obligations for AI systems. In addition, AI models and datasets typically have licenses with unclear licensing terms. SPDX profiles' capability to capture information about the declared and concluded licenses helps the practitioners with managing license compliance risk factors. SPDX 3.0 fields used in AI and Dataset Profiles are: relationshipType = hasConcludedLicense AND relationshipType = hasDeclaredLicense

**Model details:** Basic information about the model which is mandatory for AI specific applications i.e., name, version, and type. SPDX 3.0 specific fields are: spdxId, name, suppliedBy, downloadLocation, packageVersion, primaryPurpose, and releaseTime

**Model architecture:** Includes a description of the structure of the model, such as the number and type of layers, hardware characteristics and other key architectural choices. SPDX 3.0 fields are: typeOfModel, informationAboutTraining, informationAboutApplication, safetyRiskAssessment, standardCompliance, domain, and autonomyType,

**Training data and methodology:** Information about the data used to train the model, such as the size of the dataset, the data sources, and any preprocessing or data augmentation techniques used. It also includes details about the training methodology, such as the optimizer used, the loss function, and any hyperparameters that were tuned. SPDX 3.0 fields are: modelDataPreprocessing, modelExplainability, hyperparameter, energyConsumption, useSensitivePersonalInformation and relationship to training data

**Performance and verification metrics:** Information about the model's performance on various metrics, such as accuracy, precision, recall, and F1 score. It may also include information about how the model performs on different subsets of the data. For example, SPDX 3.0 includes fields for: metrics, metricDecisionThreshold

**Biases and limitations:** Lists potential biases or limitations of the model, such as imbalanced training data, overfitting, or biases in the model's predictions. It may also include information about the models limitations, such as its ability to generalize to new data or its suitability for certain use cases and compliance and risk levels. For example, SPDX 3.0 includes fields for: knownBias, limitation.

**Responsible AI considerations:** Any ethical or responsible AI considerations related to the model, such as privacy concerns, fairness, and transparency, or potential societal impacts of the model's use and standard's compliance. SPDX 3.0 fields are: knownBias, primaryPurpose

**Climate change considerations:** The main environmental concern is the energy consumption required to train and operate AI models. For example, large language models which can require significant computational power, which often translates to substantial energy use, particularly when these models are trained in data centers that rely on non-renewable energy sources. This energy use can contribute to carbon emissions, which have implications for climate change. Another consideration is the use of physical resources, such as water, for cooling data centers. We encourage developers to also log details of any energy-efficient AI algorithms, using renewable energy sources for data centers, and promoting circular economy principles for hardware in the Comments field. SPDX 3.0 fields are: energyConsumption, comments

**Dataset details:** Basic information about the dataset, i.e., name, version, license etc.), and the intended use case. SPDX 3.0 specific fields are: spdxId, name, suppliedBy, downloadLocation, packageVersion, primaryPurpose, and releaseTime

**Dataset architecture:** Includes a description of the structure of the data and type of data, such as the number SPDX 3.0 specific fields are: datasetType

**Dataset biases and limitations:** Lists potential biases or limitations of the dataset such as imbalanced training data, and production data. SPDX 3.0 specific fields are: knownBias, limitation

**Dataset responsible AI considerations:** Any ethical or responsible AI considerations related to the data, such as privacy concerns, fairness, and transparency, or potential societal impacts of the data's use. It may also include recommendations for further testing, validation, or monitoring of the model. SPDX 3.0 specific fields are: metric, knownBias, useSensitivepersonaldata and hasSensitivepersonaldata

**Key relationships used by AI and Dataset Profiles**

• relationshipType = contains

• relationshipType = hasConcludedLicense

• relationshipType = hasDeclaredLicense

• relationshipType = testedOn

• relationshipType = trainedOn

The AI Profile is designed to provide a standardized way of documenting and sharing information related to AI system and model artifacts. These artifacts are the tangible outputs of the AI development process, such as software packages, models, and datasets. It is built by layering on the Core & Software Profiles in SPDX, additional fields specific to models.

The primary challenge lies in adapting software best practices to open-source models and data, enabling the creation of automated tools. Currently, much of the essential information required for auditors, such as licensing, versioning, and verification, is missing from repositories hosting open models and data.

1. **Licensing:** Clearly stating the licenses associated with each AI artifact to ensure proper use and distribution, while respecting the rights of the creators and contributors.

2. **Versioning:** Implementing a consistent versioning scheme for AI artifacts, enabling better tracking of updates, bug fixes, and improvements over time.

3. **Verification:** Providing information on the validation and testing processes employed to assess the performance and reliability of the AI artifacts.

4. **Provenance:** Documenting the origin and history of the AI artifacts, including the creators, contributors, and any modifications made over time.

5. **Metadata:** Capturing relevant metadata related to the AI artifacts, such as the intended purpose, data use and performance, to facilitate better understanding the risk of reuse.

## Mandatory AIPackage Fields (AI Profile)

Fields in the following table are mandatory for an AIPackage to be considered conformed to an AI Profile (profileConformance = ["ai", "core", "software"]).

| Field | Cardinality | Profile | Definition |
| --- | --- | --- | --- |
| buildTime | Required(1..1) | Core | Specifies the date, time of an artifact build |
| downloadLocation | Required(1..*) | Software | Where the artifact can be found |
| name | Required(1..1) | Core | identifies the name of an artifact as designated by the creator |
| packageVersion | Required(1..1) | Software | Identify the version of an artifact |
| primaryPurpose | Required(1..1) | Software | Identify the primary purpose of the software artifact (example: "model") |
| releaseTime | Required(1..1) | Core | Specifies the date. time an artifact was released |
| spdxId | Required(1..1) | Core | globally unique identifier for the artifact |
| suppliedBy | Required(1..*) | Core | Identifies the person, organization or tool that supplied the artifact |
| relationshipType = hasConcludedLicense | Required(1..1) | Core | For every AIPackage there MUST exist exactly one Relationship of type hasConcludedLicense having that element as its from property and an AnyLicenseInfo as its "to" property. |
| relationshipType = hasDeclaredLicense | Required(1..1) | Core | For every AIPackage there MUST exist exactly one Relationship of type hasDeclaredLicense having that element as its from property and an AnyLicenseInfo as its "to" property. |

**Table 8 - Mandatory fields for AIPackage from AI Profile**

IMPLEMENTING AI BILL OF MATERIALS (AI BOM) WITH SPDX 3.0    42

## Optional AIPackage Fields (AI Profile)

| Field | Cardinality | Profile | Definition |
| --- | --- | --- | --- |
| autonomyType | Optional(0..1) | AI | List whether a human is needed in the loop of decision making |
| domain | Optional(0..1) | AI | Specifies what type of domain the AI system |
| energyConsumption | Optional(0..1) | AI | Identify the training, inference, and fine-tuning energy consumption of the AI model(s) used in an AI system. |
| energyQuantity | Optional(1..1) | AI | Specifies the energyQuantity property stores the amount of energy consumed |
| energyUnit | Optional(1..1) | AI | Specifies the energy Unit property that stores the unit used for measurement. |
| hyperparameter | Optional(0..*) | AI | Any relevant settings defined before the training process which controls the learning algorithm's behavior. |
| informationAboutTraining | Optional(0..1) | AI | Any relevant characteristics about training the model |
| informationAboutApplication | Optional(0..1) | AI | Any relevant characteristics about the application |
| limitation | Optional(0..1) | AI | Any relevant limitations known about system |
| metric | Optional(0..*) | AI | Any relevant information of how it was tested |
| metricDecisionThreshold | Optional(0..*) | AI | Any relevant information about whether or not the application reaches a benchmark |
| modelDataPreprocessing | Optional(0..*) | AI | List all steps in preparing data phase |
| modelExplainability | Optional(0..*) | AI | Any relevant information about how operating of mode can be explained to general user |
| safetyRiskAssessment | Optional(0..1) | AI | Specifies the Risk Classification of the Model |

| Field | Cardinality | Profile | Definition |
|---|---|---|---|
| standardCompliance | Optional(0..*) | AI | Any relevant standards that are complied with. |
| typeOfModel | Optional(0..*) | AI | Specifies the type of Model |
| useSensitivePersonalInformation | Optional(0..1) | AI | Does the application use any sensitive personal data |
| relationshipType = testedOn | Optional(0..1) | Core | Specifies the dataset(s) for testing purposes and deployment validation purposes |
| relationshipType = trainedOn | Optional(0..1) | Core | Specifies the dataset(s) for training purposes |

**Table 9 - Optional fields for AIPackage from AI Profile**

## Mandatory DatasetPackage Fields (Dataset Profile)

Fields in the following table are mandatory for an [DatasetPackage](#) to be considered conformed to an Dataset Profile (profileConformance = ["core", "dataset", "software"]).

| Field | Cardinality | Profile | Definition |
|---|---|---|---|
| buildTime | Required(1..1) | Core | Specifies the time and time an artifact was built |
| datasetType | Required(1..1) | Dataset | Describes the data type contained in the dataset. |
| downloadLocation | Required(1..*) | Software | Where the artifact can be found |
| originatedBy | Required(1..*) | Core | person,organization or tool that created the dataset |
| packageVersion | Required(1..1) | Software | Identify the version of an artifact |

| Field | Cardinality | Profile | Definition |
|---|---|---|---|
| primaryPurpose | Required(1..1) | Software | Identify the primary purpose of the software artifact |
| name | Required(1..1) | Core | name of artifact as designated by the creator |
| releaseTime | Required(1..1) | Core | Specifies the time and time an artifact was released |
| spdxId | Required(1..1) | Core | globally unique identifier for the artifact |
| relationshipType = hasConcludedLicense | Required(1..1) | Core | For every DatasetPackage there MUST exist exactly one Relationship of type hasConcludedLicense having that element as its from property and an AnyLicenseInfo as its "to" property. |
| relationshipType = hasDeclaredLicense | Required(1..1) | Core | For every DatasetPackage there MUST exist exactly one Relationship of type hasDeclaredLicense having that element as its from property and an AnyLicenseInfo as its "to" property. its from property and an AnyLicenseInfo as its "to" property. |

**Table 10 - Mandatory fields for DatasetPackage from Dataset Profile**

## Optional DatasetPackage Fields (Dataset Profile)

| Field | Cardinality | Profile | Definition |
|---|---|---|---|
| anonymizationMethodUsed | Optional(0..1) | Dataset | Describes the anonymization methods used. |
| confidentialityLevel | Optional(0..1) | Dataset | Describes the confidentiality level of the data points contained in the dataset. |
| dataCollectionProcess | Optional(0..1) | Dataset | Describes how the dataset was collected. |

| Field | Cardinality | Profile | Definition |
|---|---|---|---|
| dataPreprocessing | Optional(0..1) | Dataset | Describes the preprocessing steps that were applied to the raw data to create the given dataset. |
| datasetAvailability | Optional(0..1) | Dataset | Indicate if the dataset is publicly available and can be downloaded directly. Others only accessible behind a click-through, or filling a registration form. |
| datasetNoise | Optional(0..1) | Dataset | Describes what kinds of noises a dataset might have. |
| datasetSize | Optional(0..1) | Dataset | Captures how large a dataset is. |
| datasetType | Optional(0..1) | Dataset | Describes the data type contained in the dataset. |
| datasetUpdateMechanism | Optional(0..1) | Dataset | Describes a mechanism to update the dataset. |
| hasSensitivePersonalInformation | Optional(0..1) | Dataset | Describes if any sensitive personal information is present in the dataset. |
| intendedUse | Optional(0..1) | Dataset | Describes what the given dataset should be used for. |
| knownBias | Optional(0..1) | Dataset | Records the biases that the dataset is known to encompass. |
| sensor | Optional(0..1) | Dataset | Describes a sensor used for collecting the data. |

**Table 11 - Optional fields for DatasetPackage from Dataset Profile**

## Common AI and Dataset Packages Field Details

This section details fields that are common, and mandatory, to both AIPackage and DatasetPackage from AI and Dataset Profiles. Each field includes its description, type, example use cases, and a JSON-LD serialization example. Please note that all the JSON-LD examples in this entire document are for illustrative purposes only and may require additional fields to be considered a valid object within an SPDX document.

### spdxId (mandatory)

**Description:** An spdxId uniquely identifies an Element which may thereby be referenced by other Elements. These references may be internal or external. While there may be several versions of the same Element, each one needs to be able to be referred to uniquely so that relationships between Elements can be clearly articulated.

**Type:** xsd:anyURI

**Examples:** An spdxId can be used to identify Agents, Packages, Files, or any other Element. To ensure global uniqueness, an spdxId can incorporate a standardized unique identifier, such as a Universally Unique Identifier (UUID).

**Syntax:**

```
{
    "type": "ai_AIPackage",
    "spdxId": "https://spdx.org/spdxdocs/Person/
    AS-1000e6a2-0229-4875-baa7-c99be213b6e1"
}
```

### name (mandatory)

**Description:** Identifies the name of an Element as designated by the creator.

**Type:** xsd:string

**Examples:** A name field can be used to identify various elements within an SPDX document, including the document itself, a Package, a File, or a specific code snippet within a File.

**Syntax:**

```
{
    "type": "ai_AIPackage",
    "name": "An example SPDX document"
}
```

### buildTime (mandatory)

**Description:** Specifies the time an artifact was built. buildTime is a string representation of specific date and time of a build time. It has a resolution of seconds and is always expressed in UTC time zone. Its ISO-8601 format is YYYY-MM-DDThh:mm:ssZ.

**Type:** DateTime (subclass of xsd:dateTimeStamp)

**Examples:** Package, File, Dataset, Artifact build time.

**Syntax:**

```
{
    "type": "ai_AIPackage",
    "buildTime": "2024-04-24T12:00:00Z"
}
```

### downloadLocation (mandatory)

**Description:** A downloadLocation identifies the download Uniform Resource Identifier for the package at the time that the document was created. Where and how to download the exact package being referenced is critical for verification and tracking data.

**Type:** xsd:anyURI

**Examples:** The downloadLocation can point to different types of resources, including direct download links, repository URLs, or even specific paths within a version control system.

**Syntax:**

```
{
    "type": "ai_AIPackage",
    "downloadLocation": "https://example.com/
    download/anotherexamplepackage.tar.gz"
}
```

## packageVersion (mandatory)

**Description:** A packageVersion is useful for identification purposes and for indicating later changes of the package version. There are no restrictions on the versioning scheme used.

**Type:** xsd:string

**Examples:** Valid examples include "3.14159", "1.0.0-alpha", "2.4_13", "24.04", "1.2.1.2", "2024H", "961219 ASIA", and "3.6:1:0123abcd:x86_64".

**Syntax:**

```
{
    "type": "ai_AIPackage",
    "packageVersion": "3.14159"
}
```

## primaryPurpose (mandatory)

**Description:** primaryPurpose provides information about the primary intended function of the Software Artifact. Its value must be selected from one of the entries defined in the SoftwarePurpose data type.

**Type:** SoftwarePurpose (select one from this list: application, archive, bom, configuration, container, data, device, deviceDriver, diskImage, documentation, evidence, executable, file, filesystemImage, firmware, framework, install, library, manifest, model, module, operatingSystem, other, patch, platform, requirement, source, specification, test)

**Examples:** "application" indicates that the Software Artifact is a software application. "library" indicates that it is a software library. "model" indicates that it is a machine learning or artificial intelligence model.

**Syntax:**

```
{
    "type": "ai_AIPackage",
    "primaryPurpose": "model"
}
```

## releaseTime (mandatory)

**Description:** Specifies the time an artifact was released. releaseTime is a string representation of specific date and time of a release time. It has a resolution of seconds and is always expressed in UTC time zone. Its ISO-8601 format is YYYY-MM-DDThh:mm:ssZ.

**Type:** DateTime (subclass of xsd:dateTimeStamp)

**Examples:** Release time for Packages, Files, Datasets, and Artifacts.

**Syntax:**

```
{
    "type": "dataset_DatasetPackage",
    "releaseTime": "2023-10-06T17:00:00Z"
}
```

## suppliedBy (mandatory)

**Description:** identify the actual distribution source for the artifact (e.g., snippet, file,AI, Dataset package, vulnerability

**Type:** Agent (Organization or Person)

**Examples:** The suppliedBy field helps in tracking the suppliers of software components, making it easier to identify the origin of each component and ensures that the correct suppliers are credited, which is crucial for compliance with licensing terms and auditing purposes.

**Syntax:**

```
{
    "type": "ai_AIPackage",
    "suppliedBy": {
        "type": "Organization",
        "name": "Example AI Co-op"
                }
}
```

relationshipType = hasConcludedLicense (mandatory)

**Description:** The hasConcludedLicense relationship identifies the license that the SPDX data creator has come to a reasonable conclusion that it governs the Software Artifact. This conclusion is based on an analysis of the license information within the Software Artifact and other relevant data. A Software Artifact, like AI Package and Dataset Package, can have multiple concluded licenses.

**Type:** Relationship

**Examples:** The license in the "to" field within a Relationship object can be any object of AnyLicenseInfo class, including NoAssertionLicense, NoneLicense, LicenseExpression, or any license listed in the SPDX License List.

**Syntax:**

```
{
    "type": "Relationship",
    "relationshipType": "hasConcludedLicense",
    "from": "https://spdx.org/spdxdocs/
AIPackage/EX-a09c4e3e-df9a-48e7-9a2a-
38ca15cd2ea7",
    "to": [
        "https://spdx.org/licenses/Apache-2.0"
]
}
```

relationshipType = hasDeclaredLicense (mandatory)

**Description:** The hasDeclaredLicense relationship identifies the license information actually found in the Software Artifact, for example as detected by use of automated tooling. A Software Artifact, like AI Package and Dataset Package, can have multiple concluded licenses.

**Type:** Relationship

**Examples:** The license in the "to" field within a Relationship object can be any object of AnyLicenseInfo class, including NoAssertionLicense, NoneLicense, LicenseExpression, or any license listed in the SPDX License List.

**Syntax:**

```
{
    "type": "Relationship",
    "relationshipType": "hasDeclaredLicense",
    "from": "https://spdx.org/spdxdocs/
DatasetPackage/DS-d170dabb-fe05-4c98-
b41d-5f62dc6d606a", "to":
    [
      "https://spdx.org/licenses/CC-BY-4.0"
    ]
}
```

## Specific AIPackage Field Details

This section details fields that are specific to the AIPackage from the AI Profile. Each field includes its description, type, example use cases, and a JSON-LD serialization example.

### autonomyType (optional)

**Description:** Indicates whether the system can perform a decision or action without human involvement or guidance.

**Type:** PresenceType (select one from this list: yes, no, noAssertion)

**Examples:** "yes" indicates that AI systems potentially have no humans in the loop e.g. self-driving cars, autonomous Aerial Vehicles, robots, virtual assistants.

"no" indicates that humans are accountable and in the loop to validate AI decision making e.g. mortgage loan approval, financial investing and justice system.

"noAssertion" indicates that the autonomy type is unclear from the information available, when the model is being documented.

**Syntax:**

```
{
    "type": "ai_AIPackage",
    "ai_autonomyType": "yes"
}
```

### domain (optional)

**Description:** A free-form text that describes the domain where the AI model contained in the AI software can be expected to operate successfully.

**Type:** xsd:string

**Examples:** computer vision, natural language processing, etc.

**Syntax:**

```
{
    "type": "ai_AIPackage",
    "ai_domain": "natural language processing"
}
```

### energyConsumption (optional)

**Description:** Captures known or estimated energy consumption for the training of the AI model. In case it's not known, the estimation could be based on information about computational resources used (e.g. number of floating point operations), training time, type and quantity of processing units, and other relevant details related to the training. If energyConumption has a value, then energyQuantity and engergyUnit are mandatory.

**Type:** EnergyConsumption

**Examples:** The energy consumption of an NLP type large language model) type application, used in machine translation, text generation, and sentiment analysis. requires vast amounts of computational resources to train. The energy consumption can be significant,and it is estimated that a single large language model can emit as much carbon as five cars over their entire lifetimes. As a result, there is growing interest in developing more energy-efficient methods for training NLP models, but these fields will start the ability for developers to start optimizing training algorithms, and leveraging more efficient hardware.

**Syntax:**

```
{
    "type": "ai_AIPackage",
    "ai_energyConsumption":
        {
            "type": "ai_EnergyConsumption",
            "ai_trainingEnergyConsumption":
            [
                {
                    "type": "ai_
                    EnergyConsumptionDescription",
                    "ai_energyQuantity": "36.5",
                    "ai_energyUnit":
                    "kilowattHour"
                },
                {
                    "type": "ai_
                    EnergyConsumptionDescription",
                    "ai_energyQuantity": "0.4",
                    "ai_energyUnit":
                    "kilowattHour"
                }
            ],
            "ai_inferenceEnergyConsumption":
            [
                {
                    "type": "ai_
                    EnergyConsumptionDescription",
                    "ai_energyQuantity": "0.042",
                    "ai_energyUnit":
                    "kilowattHour"
                }
            ]
        }
}
```

### energyQuantity (optional)

**Description:** Provides the quantity information of the energy.

**Type:** xsd:decimal

**Examples:** power usage of each sensor such as traffic patterns.

**Syntax:**

```
{
    "type": "ai_EnergyConsumptionDescription",
    "ai_energyQuantity": "0.042",
    "ai_energyUnit": "kilowattHour"
}
```

### energyUnit (optional)

**Description:** Provides the unit information of the energy.

**Type:** EnergyUnitType (select one from this list: kilowattHour, megajoule, other)

**Examples:** "kilowattHour" for Kilowatt-hour (kW.h). "megajoule" for Megajoule (MJ). "other" for any other units of energy measurement (use "comment" field to specify).

**Syntax:**

```
{
    "type": "ai_EnergyConsumptionDescription",
    "ai_energyQuantity": "0.042",
    "ai_energyUnit": "kilowattHour"
}
```

## finetuningEnergyConsumption (optional)

**Description:** specifies the amount of energy consumed when fine tuning the AI model that is being used in the AI system.

**Type:** EnergyConsumptionDescription

**Examples:** During the fine-tuning stage of a deep learning neural network for image classification, the fine-tuning module itself is reported to have consumed 15 kWh (kilowatt-hour) of energy.

**Syntax:**

```
{
    "type": "ai_EnergyConsumption",
    "ai_finetuningEnergyConsumption":
    [
        {
            "type": "ai_
            EnergyConsumptionDescription",
            "ai_energyQuantity": "2.4",
            "ai_energyUnit": "kilowattHour"
        }
    ]
}
```

## hyperparameter (optional)

**Description:** Records a hyperparameter value. Hyperparameters are settings defined before the training process that control the learning algorithm's behavior. They differ from model parameters, which are learned from the data during training. Developers typically set hyperparameters manually or through a process of hyperparameter tuning (also known as trial and error).

**Type:** /Core/DictionaryEntry

**Examples:** learning rate, batch size, and the number of layers in a neural network.

**Syntax:**

```
{
    "type": "ai_AIPackage",
    "ai_hyperparameter":
        [
        {
            "type": "DictionaryEntry",
            "key": "cnn_kernel_vals",
            "value": "[5, 5, 3, 3, 3]"
        },
        {
            "type": "DictionaryEntry",
            "key": "beam_search_scoring_
            mode", "value": "Words"
        }
        ]
}
```

## inferenceEnergyConsumption (optional)

**Description:** specifies the amount of energy consumed during inference time by an AI model that is being used in the AI system.

**Type:** EnergyConsumptionDescription

**Examples:** During the inference stage of a deep learning neural network for image classification, the training module itself is reported to have consumed 80 watt-hour of energy.

**Syntax:**

```
{
    "type": "ai_EnergyConsumption",
    "ai_inferenceEnergyConsumption": [
        {
```

```
            "type": "ai_
            EnergyConsumptionDescription",
            "ai_energyQuantity": "0.042",
            "ai_energyUnit": "kilowattHour"
        }
    ]
}
```

### informationAboutApplication (optional)

**Description:** Description of how the AI model is used within the software. It should include any relevant information, such as pre-processing steps, third-party APIs, and other pertinent details. Functionality provided by the AI model within the software application, including: any specific tasks or decisions it is designed to perform; any pre-processing steps that are applied to the input data before it is fed into the AI model for inference, such as data cleaning, normalization, or feature extraction; and any third-party APIs or services that are used in conjunction with the AI model, such as data sources, cloud services, or other AI models. Description of any dependencies or requirements needed to run the AI model within the software application, including: specific hardware, software libraries, third-party APIs and operating systems and other pertinent details.

**Type:** xsd:string

**Examples:** Any specific tasks or decisions that were made in design, develop and deployment of this model.

**Syntax:**

```
{
    "type": "ai_AIPackage",
    "ai_informationAboutApplication": "A vehicle identification system utilizes XYZ Cloud's object classification service in conjunction with a custom-AI model designed for vehicle make and model classification. The system is designed to process 1600x1200 pixel images captured by a consumer-grade camera equipped with automatic lighting adjustment."
}
```

### informationAboutTraining (optional)

**Description:** A detailed explanation of the training process, including the specific techniques, algorithms, and methods employed.

**Type:** xsd:string

**Examples:** Any relevant information about the training data used to train the AI model, such as the source, quality with the measurements/benchmarks used. A specific example is to describe the training process for a sentiment analysis model designed to classify text as positive, negative, or neutral.

**Syntax:**

```
{
    "type": "ai_AIPackage",
    "ai_informationAboutTraining": "The sentiment analysis model was trained using a supervised learning approach with the following details: The training data was sourced from a combination of public datasets such as the IMDb movie reviews dataset and the Sentiment140 dataset. The data was preprocessed to remove duplicates, handle missing values, and normalize text. The quality was measured using metrics such as accuracy, precision, recall, and F1 score
```

on a held-out validation set.The model was trained using a deep learning approach with a Bidirectional LSTM (Long Short-Term Memory) network. The network architecture included an embedding layer, two bidirectional LSTM layers, and a dense output layer with softmax activation. The model was optimized using the Adam optimizer with a learning rate of 0.001. The model achieved an accuracy of 85% on the test set, with an F1 score of 0.84 for positive sentiment, 0.82 for negative sentiment, and 0.80 for neutral sentiment."

}

limitation (optional)

**Description:** Captures a limitation of the AI package (or of the AI models present in the AI package). Note that this is not guaranteed to be exhaustive and captures a limitation of the AI package (or of the AI models present in the AI package). Note that this is not guaranteed to be exhaustive.

**Type:** xsd:string

**Examples:** Regional restrictions on how a model or application can be used. Poor data diversity in training datasets, such as a predominance of images featuring light-skinned individuals. Algorithm limitations, such as sensitivity to noise. Domain-specific limitations, such as the distinct challenges involved in training a model to detect credit card fraud compared to detecting fraud in other financial transactions.

**Syntax:**

```
{
    "type": "ai_AIPackage",
```

"ai_limitation": "The dataset used for model training was largely collected under clear weather conditions, which may limit the model's ability to predict accurately in other weather types."

}

metric (optional)

**Description:** Records the measurement with which the AI model was evaluated. This makes statements about the prediction quality including uncertainty, accuracy, characteristics of the tested population, quality, fairness, explainability, robustness etc.

**Type:** DictionaryEntry

**Examples:** accuracy, precision, recall, F1-score, ROC curve, AUC, MSE, MAE, RMSE, R-squared, list benchmark, personas.

**Syntax:**

```
{
    "type": "ai_AIPackage",
    "ai_metric": [
        {
            "type": "DictionaryEntry",
            "key": "precision",
            "value": "0.94"
        },
        {
            "type": "DictionaryEntry",
            "key": "F1",
            "value": "0.91"
        }
    ]
}
```

metricDecisionThreshold (optional)

**Description:** Captures the threshold that was used for computation of a metric described in the metric field. Predefined threshold value used to make decisions based on the output of a predictive model. The threshold is applied to the model's output score or probability to determine the predicted class or category for a given input. The choice of the metricDecisionThreshold value can have a significant impact on the performance of the AI system. A low threshold may result in a higher number of false positives, while a high threshold may result in a higher number of false negatives. Therefore, it is essential to carefully select an appropriate threshold value based on the specific requirements and trade-offs of the application.

**Type:** DictionaryEntry

**Examples:** The value used to determine whether the input should be classified as positive or negative. If the probability score is greater than the threshold, the input is classified as positive; otherwise, it is classified as negative.

**Syntax:**

```
{
    "type": "ai_AIPackage",
    "ai_metricDecisionThreshold": [
        {
            "type": "DictionaryEntry",
            "key": "precision",
            "value": "0.20"
        }
    ]
}
```

modelDataPreprocessing (optional)

**Description:** Set of operations performed on the input data before it is fed into a machine learning model for training or inference. The goal of modelDataPreprocessing is to transform the raw data into a suitable format that can be used by the model to improve its accuracy, efficiency, and robustness.

**Type:** xsd:string

**Examples:** Data cleaning, Data normalization, Data transformation, Data splitting, Data augmentation:

**Syntax:**

```
{
    "type": "ai_AIPackage", "ai_
modelDataPreprocessing": "lower casing all
text, punctuation marks removed, text shorter
than 10 characters removed, leave-one-out
cross-validation applied"
}
```

modelExplainability (optional)

**Description:** A free-form text that lists the different explainability mechanisms and how they can be used to explain the results from the AI model. This field promotes insights into how the model arrived at its predictions, decisions, or recommendations, and why it made those choices. The goal of model explainability is to make AI systems more transparent, trustworthy, and accountable.

**Type:** xsd:string

**Examples:** Feature importance, partial dependence plots, input features and predictions relationship visualization, SHAP (SHapley Additive exPlanations), LIME (Local Interpretable Model-Agnostic Explanations), rule-based explanations. While it is preferred to use standardized method names, there is no restriction on what the value can be. More details of how the method has been applied to the model can additionally be placed in the "description" field.

Syntax:

```
{
    "type": "ai_AIPackage",
    "ai_modelExplainability": "feature_importance",
    "description": "The AI package utilizes a random forest model for image classification. Feature importance is calculated using permutation importance to determine the most influential pixels in the images."
}
```

safetyRiskAssessment (optional)

**Description:** Records the results of general safety risk assessment of the AI system.

Using categorization according to the EU general risk assessment methodology. The methodology implements Article 20 of Regulation (EC) No 765/2008 and is intended to assist authorities when they assess general product safety compliance. US FDA uses these definitions as well, but note : that this categorization differs from the one proposed in the EU AI Act's provisional agreement.

**Type:** SafetyRiskAssessmentType (select one from this list: serious, high, medium, low)

**Examples:** serious (eg. self-driving cars), high (eg. adaptive education), medium (eg. bank loan approval), low (eg. shopping recommendation). The developer would categorize their application's safetyRiskassessment per the intended audience such as application sold in EU.

Syntax:

```
{
    "type": "ai_AIPackage",
    "ai_safetyRiskAssessment": "serious"
}
```

standardCompliance (optional)

**Description:** Captures a standard that the AI software complies with. This includes both published and unpublished standards, such as those developed by ISO, IEEE, and ETSI. The standards may, but are not necessarily required to, satisfy a legal or regulatory requirement. There is a separated field "standardName", in the Core Profile, where developers can log other standards adhered to but that compliance was not obtained.

**Type:** xsd:string

**Examples:** Any relevant standard from DIN, ETSI, IEC, IEEE, ISO, ITU, JISC, NIST, OASIS, W3C, etc. e.g. ISO/IEC 5962:2021, ISO/IEC TS 4213:2022, IEEE 7014-2024, FGAI4AD-02.

Syntax:

```
{
    "type": "ai_AIPackage",
    "ai_standardCompliance": "IEEE 7002-2022 Data Privacy Processing"
}
```

trainingEnergyConsumption (optional)

**Description:** Specifies the amount of energy consumed when training the AI model that is being used in the AI system.

**Type:** EnergyConsumptionDescription

IMPLEMENTING AI BILL OF MATERIALS (AI BOM) WITH SPDX 3.0    56

**Examples:** During the training stage of a deep learning neural network for image classification, the training module itself is reported to have consumed 980 kWh (kilowatt-hour) of energy.

**Syntax:**

```
{
    "type": "ai_EnergyConsumption",
    "ai_trainingEnergyConsumption":
    [
        {
            "type": "ai_
            EnergyConsumptionDescription",
            "ai_energyQuantity": "980",
            "ai_energyUnit": "kilowattHour"
        }
    ]
}
```

typeOfModel (optional)

**Description:** Records the type of the AI model(s) used in the software.

**Type:** xsd:string

**Examples:** Supervised model, unsupervised model, reinforcement learning model or a combination of those. Or neural network, linear model, support vector machines and Bayesian models.

**Syntax:**

```
{
    "type": "ai_AIPackage",
    "ai_typeOfModel": "reinforcement
    learning"
}
```

useSensitivePersonalInformation (optional)

**Description:** If personal data is used in a context-specific and purpose-limited manner, taking into account the user's preferences, expectations, and consent. Using sensitive personal information is often used to personalize or adapt the AI system's behavior to the user's needs, preferences, or context.

**Type:** PresenceType (select one from this list: yes, no, noAssertion)

**Examples:** If sensitive personal data likeLocation data, Health data, Biometric data, Behavioral data, etc. are used, then value should be 'yes' otherwise 'no'

**Syntax:**

```
{
    "type": "ai_AIPackage",
    "ai_useSensitvePersonalInformation": "yes"
}
```

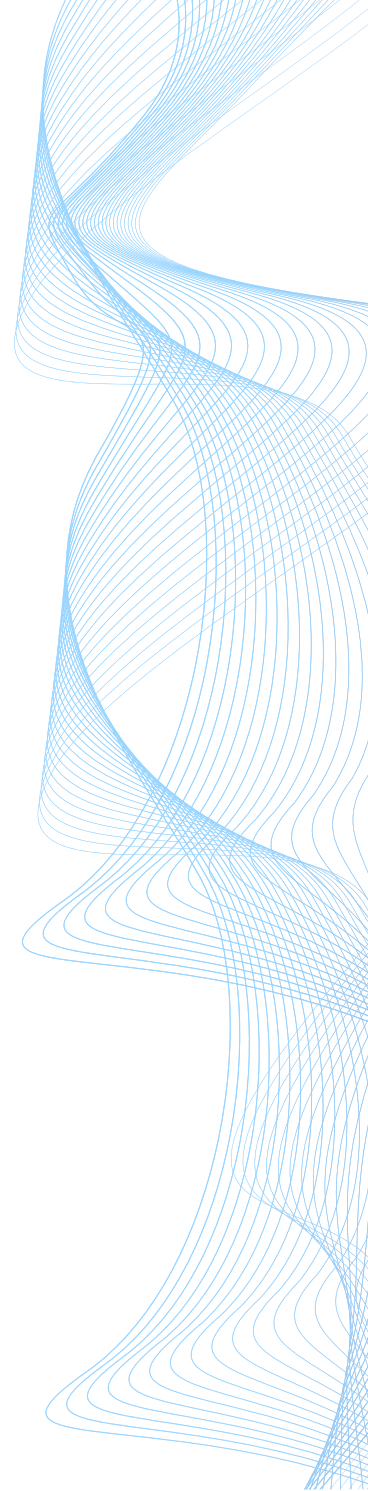

## Specific DatasetPackage Field Details

This section details fields that are specific to the DatasetPackage from the Dataset Profile. Each field includes its description, type, example use cases, and a JSON-LD serialization example.

### anonymizationMethodUsed (optional)

**Description:** A free-form text that describes the methods used to anonymize the dataset or fields in the dataset.

**Type:** xsd:string

**Examples:** Pseudonymization, k-anonymity, l-diversity, t-closeness, differential privacy, and other methods. While it is preferred to use standardized method names, there is no restriction on what the value can be. More details of how the method has been applied to the dataset, including any anonymization-related pre-processing steps, can additionally be placed in the "description" field.

**Syntax:**

```
{
    "type": "dataset_DatasetPackage",
    "dataset_anonymizationMethodUsed":
    "pseudonymization",
    "description": "replace direct identifiers
(such as name or social security number)
with artificial identifiers to prevent the
data from being directly linked back to the
individual"
}
```

### confidentialityLevel (optional)

**Description:** Describes the different confidentiality levels as given by the Traffic Light Protocol.

**Type:** ConfidentialityLevelType (select one from this list: red, amber, green, clear)

- red: Data points in the dataset are highly confidential and can only be shared with named recipients
- amber: Data points in the dataset can be shared only with specific organizations and their clients on a need to know basis
- green: Dataset can be shared within a community of peers and partners
- clear: Dataset may be distributed freely, without restriction.

**Examples:** A dataset could be marked with "red" if it includes sensitive financial information. The "amber" dataset could include proprietary business data, customer lists, or competitive intelligence. The "green" dataset might include general business data. The "clear" dataset might include publicly available data.

**Syntax:**

```
{
    "type": "dataset_DatasetPackage",
    "dataset_confidentialityLevel": "clear"
}
```

### dataCollectionProcess (optional)

**Description:** Describes how the dataset was collected (including all sources).

**Type:** xsd:string

**Examples:** Examples include the sources from which a dataset was scrapped and the interview protocol that was used for data collection. The field can also record if the dataset is a subset from another dataset or if it was created by combining multiple datasets.

**Syntax:**

```
{
    "type": "dataset_DatasetPackage",
    "dataset_dataCollectionProcess":
"Collected by scraping data from https://
example.com"
}
```

### dataPreprocessing (optional)

**Description:** Describes the various preprocessing steps that were applied to the raw data to create the dataset.

**Type:** xsd:string

**Examples:** Standardization, normalization, deduplication, tokenization, and any pre-processing steps that are applied to the input data before it is fed into the AI model such as data cleaning, normalization, or feature extraction, etc.

**Syntax:**

```
{
    "type": "dataset_DatasetPackage",
    "dataset_dataPreprocessing": "z-score
    standardization",
    "description": "each data point is re-
    scaled based on the mean and standard
    deviation of the dataset."
}
```

### datasetAvailability (optional)

**Description:** Some datasets are publicly available and can be downloaded directly. Others are only accessible behind a clickthrough, or after filling a registration form. This field will describe the dataset availability from that perspective.

**Type:** DatasetAvailabilityType (select one from this list: clickthrough, directDownload, query, registration, scrapingScript)

- **clickthrough**: the dataset is not publicly available and can only be accessed after affirmatively accepting terms on a clickthrough webpage,

- **directDownload**: the dataset is publicly available and can be downloaded directly,

- **query**: the dataset is publicly available, but not all at once, and can only be accessed through queries which return parts of the dataset,

- **registration**: the dataset is not publicly available and an email registration is required before accessing the dataset, although without an affirmative acceptance of terms,

- **scrapingScript**: the dataset provider is not making available the underlying data and the dataset must be reassembled, typically using the provided script for scraping the data.

**Examples:** Examples of how allowing users to understand how the dataset can be accessed that is clickthrough type is OpenStreetMap (OSM) which is a collaborative project to create a free editable map of the world. The OSM data includes geographical features such as roads, buildings, and points of interest. To download the data, users must visit the OSM website and agree to the terms of use.

**Syntax:**

```
{
    "type": "dataset_DatasetPackage",
    "dataset_datasetAvailability": "
    clickthrough" "
}
```

### datasetNoise (optional)

**Description:** Describes what kinds of noises a dataset might encompass.

**Type:** xsd:string

**Examples:** Inconsistent data refers to data that is not uniform or standardized. Missing data refers to data points that are not present in the dataset, Irrelevant data refers to data that is not relevant to the problem being solved by the AI model and human error can also introduce noise into a dataset.

**Syntax:**

```
{
    "type": "dataset_DatasetPackage",
    "dataset_datasetNoise": "Human error.
    Since manually entered into the system,
    errors such as typos or incorrect data
    entry can occur."
}
```

### datasetSize (optional)

**Description:** Captures how large a dataset is, in bytes. The size is to be measured in bytes.

**Type:** xsd:nonNegativeInteger

**Examples:** Dataset sizes vary and this field captures the actual bytes.

**Syntax:**

```
{
    "type": "dataset_DatasetPackage",
    "dataset_datasetSize": 2689
}
```

### datasetType (mandatory)

**Description:** Specifies the data types contained within the dataset. A dataset may include multiple data types.

**Type:** DatasetType (select one from this list: audio, categorical, graph, image, noAssertion, numeric, other, sensor, structured, syntactic, text, timeseries, timestamp, video)

**Examples:** "structured" for the data organized in tabular format or retrieved from a relational database. "timestamp" for the data that includes a timestamp for each entry, but may not be ordered or recorded at specific intervals. "noAssertion" for which the data type cannot be determined.

**Syntax:**

```
{
    "type": "dataset_DatasetPackage",
    "dataset_datasetType":
    [
        "structured",
        "timestamp"
    ]
}
```

### datasetUpdateMechanism (optional)

**Description:** A free-form text that describes a mechanism to update the dataset.

**Type:** xsd:string

**Examples:** Batch, real-time (such as stock prices), incremental (incremental updates such as social media data or news articles), and manual (such as data that require human judgment to verify).

**Syntax:**

```
{
    "type": "dataset_DatasetPackage",
    "dataset_datasetUpdateMechanism":
    "Batch. Updated annually."
}
```

### hasSensitivePersonalInformation (optional)

**Description:** Indicates the presence of sensitive personal data or information that allows drawing conclusions about a person's identity.

**Type:** PresenceType (select one from this list: yes, no, noAssertion)

**Examples:** "yes" indicates that personal sensitive information is being used like biometrics data. Information about an individual's race or ethnicity. "No" indicates that no personal sensitive information is used in training, testing or productions.

**Syntax:**

```
{
    "type": "dataset_DatasetPackage",
    "dataset_hasSensitivePersonalInformation":
    "no"
}
```

### intendedUse (optional)

**Description:** A free-form text that describes what the given dataset should be used for. Using a dataset for a purpose other than its intended purpose can lead to vulnerabilities and legal issues.

**Type:** xsd:string

**Examples:** If a dataset is collected for research purposes, using it for commercial purposes may violate intellectual property rights or other legal agreements When medical data collected from a specific demography might only be applicable for training machine learning models to make predictions for that demography. In such a case, the intendedUse field would capture this information.

**Syntax:**

```
{
    "type": "dataset_DatasetPackage",
    "dataset_intendedUse": "To make the
    research about greenhouse gas emissions
    accessible."
}
```

### knownBias (optional)

**Description:** A free-form text that describes the different biases that the dataset encompasses.

**Type:** xsd:string

**Examples:** i) Selection bias: Occurs when the data collected is not representative of the population being studied. ii) Measurement bias: For example, if a survey question is worded in a way that influences the response, or if certain measurements are more accurate for some groups than others. iii) Label bias: Occurs when the labels used to categorize the data are biased in some way. i.e. if certain labels are more commonly applied to certain groups than others. etc.

**Syntax:**

```
{
    "type": "dataset_DatasetPackage",
    "dataset_knownBias": "Data in some
    geographical areas are more complete
    than the others."
}
```

### sensor (optional)

**Description:** Describes sensors used for data collection, including its calibration value. Values are stored in a key-value format.

**Type:** DictionaryEntry

**Examples:** Keys can include camera, lidar, radar, microphone, temperature, pressure, proximity, biometric sensors, or other relevant types. Values should ideally be standardized model and manufacturer codes, if available. However, there are no restrictions on the specific keys or values used.

**Syntax:**

```
{
    "type": "dataset_DatasetPackage",
    "dataset_sensor":
    [
        {
            "type": "DictionaryEntry",
            "key": "lidar",
            "value": "Acme A-5.2M"
        },
        {
            "type": "DictionaryEntry",
            "key": "lidar-calibration-
            distance-offset",
            "value": "0.05"
        }
    ]
}
```

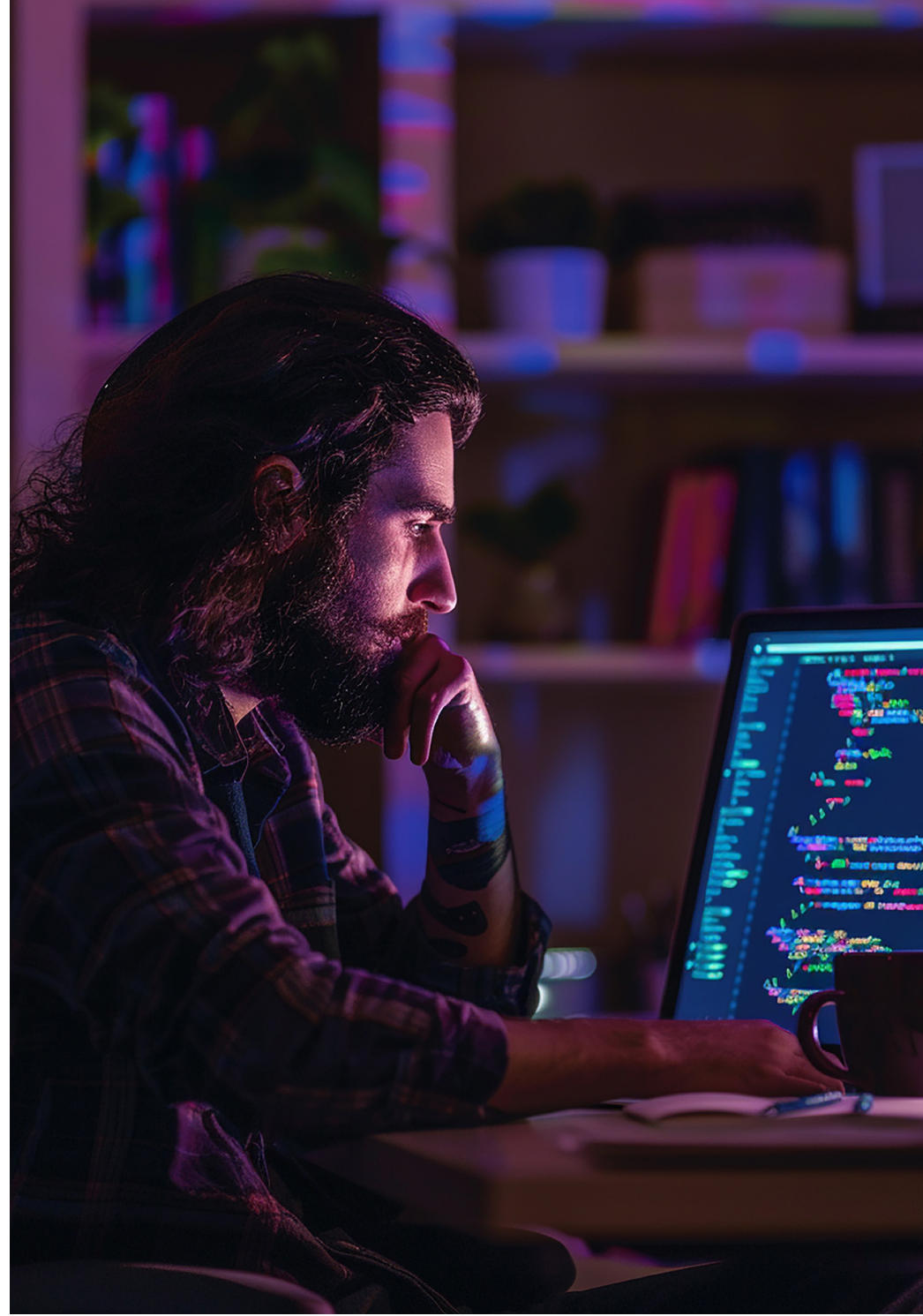

# Real World Evidence Examples

Here are a few examples of different types of AI systems with the corresponding SPDX 3.0 files. A comprehensive collection of examples, including those to be added in the future, is available at: https://github.com/spdx/spdx-examples.

## Handwritten Text Recognition Application (SimpleHTR)

The complexity of an AI BOM can increase significantly even for moderately complex AI systems for several reasons. Firstly, AI systems often have numerous dependencies, especially when utilizing popular open-source frameworks like TensorFlow. Secondly, as the complexity of the AI model increases, such as using a model like OpenAI's GPT-4, the representation becomes increasingly verbose, which can be challenging for readers to follow. While it is entirely possible to create AI BOMs for highly complex systems, using such examples would defeat the purpose by impacting readability and understandability.

To provide a simple yet non-trivial real-world example of AI BOM, we decided to select a publicly available open-source AI system with a manageable amount of dependencies. To ensure a high level of quality, we limited our consideration to projects with over 1,000 GitHub stars. After thorough analysis and deliberation among the paper's authors, we selected the SimpleHTR project for the demonstration purpose.

### SimpleHTR Overview

SimpleHTR is a handwritten text recognition system developed by Harald Scheidl. Implemented using TensorFlow, SimpleHTR is trained on the IAM handwritten text recognition dataset. The system's AI model can process images of single handwritten words or multiple lines of handwritten words, outputting the recognized text. Both the line-level and word-level models are neural network-based, comprising five CNN layers and two RNN (LSTM) layers, utilizing a connectionist temporal loss and decoding method.

SimpleHTR provides the source code for model training, data loading, and data preprocessing, employing various open-source packages to achieve these functionalities. For more detailed information about the model, dataset preprocessing, and inference process, please refer to the developer's blog post [t].

### SPDX System BOM which includes the AI Profile for SimpleHTR

The AI BOM for SimpleHTR captures all relevant information about the AI models, the datasets used for training, the packages supporting these functionalities, their dependencies, and the license information of these packages. It also includes information about the provenance of the AI BOM document itself.

For example, information about a word model for SimpleHTR is captured inside an AIPackage and information about the IAM Handwriting Database is captured inside a DatasetPackage. A Relationship instance with relationshipType: "trainedOn" defines a semantic link between these two Packages.

AIPackage(name="word-model")

      trainedOn DatasetPackage(name="IAMdataset")

The complete AI BOM is available in our official example repositories. While a detailed instruction manual on creating the AI BOM is beyond the scope of this paper, we plan to make such a tutorial available in the near future.

## CO2 Dataset

We selected the "Data on CO2 and Greenhouse Gas Emissions" dataset (CO2 dataset) from Our World in Data to demonstrate a real-world example of a dataset BOM for an existing dataset. This choice is driven by two key factors: the dataset's structural simplicity and the accessibility of its domain to the general public (news outlets have consistently reported on greenhouse gas emissions for a long time). The entire CO2 dataset consists of just two files (data and codebook) and is freely available at https://github.com/owid/co2-data/.

Despite its simple structure, the dataset's underlying data is sourced from various origins, leading to potential complexities like varying licensing. For this document, we'll focus on specific fields and relationships. The complete dataset BOM is available at https://github.com/spdx/spdx-examples/tree/master/dataset/example01

The dataset consists of two plain text files: data.csv and codebook.csv, both are in CSV (comma-separated values) format. data.csv contains yearly emission data for countries, with 80 columns defined in a header. Most data is numerical (e.g., population, GDP, CO2 emissions), with some categorical (e.g., country). codebook.csv details the columns in data.csv, including descriptions, units, and data sources.

In the BOM, data.csv and codebook.csv are defined as File instances with primaryPurpose: "data" and contentType: "text/csv;charset=UTF-8". A Relationship instance with relationshipType: "describes" defines a semantic link between these two Files.

`codebook.csv describes data.csv`

The rootElement of the BOM is a DatasetPackage instance. A relationship with relationshipType "contains" is used to link data.csv and codebook.csv Files with the DatasetPackage instance.

`DatasetPackage1 contains [data.csv, codebook.csv]`

A Relationship instance with relationshipType: hasDeclaredLicense is used to describe the license of the dataset package.

`DatasetPackage1 hasDeclaredLicense CC-BY-4.0`

The DatasetPackage class has a number of properties to describe a dataset's characteristics. For example, hasSensitivePersonalInformation is set to "no," and knownBias is set to the string "Data in some geographical areas are more complete than others."

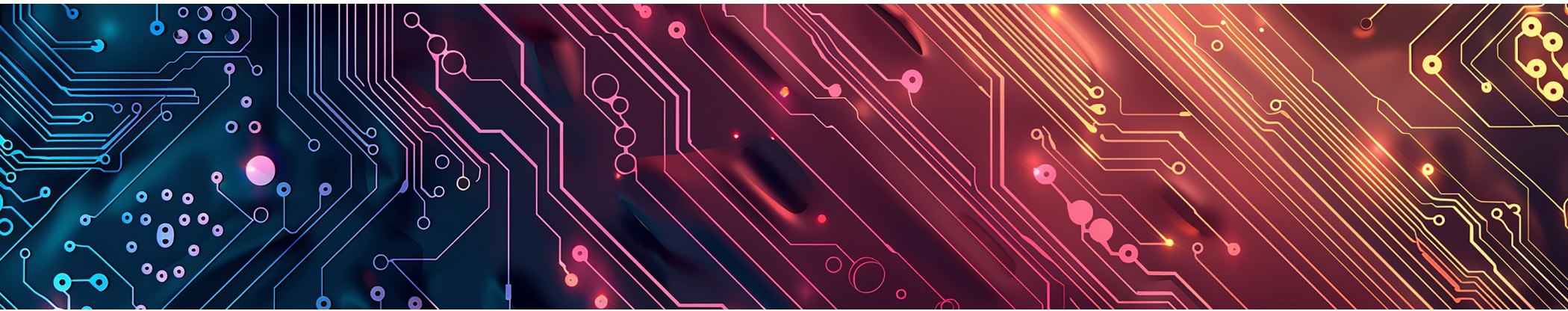

# Future Directions

Systems today are increasingly incorporating trained AI/ML models into products. These profiles are the starting point to capturing relevant metadata in a way that can be incorporated into system analysis so that risk can be assessed. However, for applications with safety critical other "ingredients", may also need to be linked into an effective knowledge graph, so that analysis can be applied.

The evolution of SPDX is poised to transform system analysis and risk assessment by broadening its scope to encompass a wider array of components. Leveraging the established SPDX foundational core model, which features common element and artifact classes, the initiative is prepared to introduce additional profiles for hardware, services, testing, behavior analysis, and operations. This comprehensive strategy will facilitate intricate linkages between software, hardware, models, and services, thereby enabling robust system safety analysis and a nuanced understanding of system behavior and context.

Moreover, the development of profiles for operations and threat/harm will augment the SPDX toolkit, enhancing its capabilities for assessing and mitigating risks in complex systems. Looking forward, SPDX aims to achieve further alignment with ISO and IEEE standards, as well as validate compliance with worldwide government acts, to ensure transparency and accountability through meticulous documentation.

This strategic direction will not only fortify the reliability and safety of AI/ML-integrated systems but also pave the way for more holistic and effective risk management practices across industries and other documented risk scenarios.

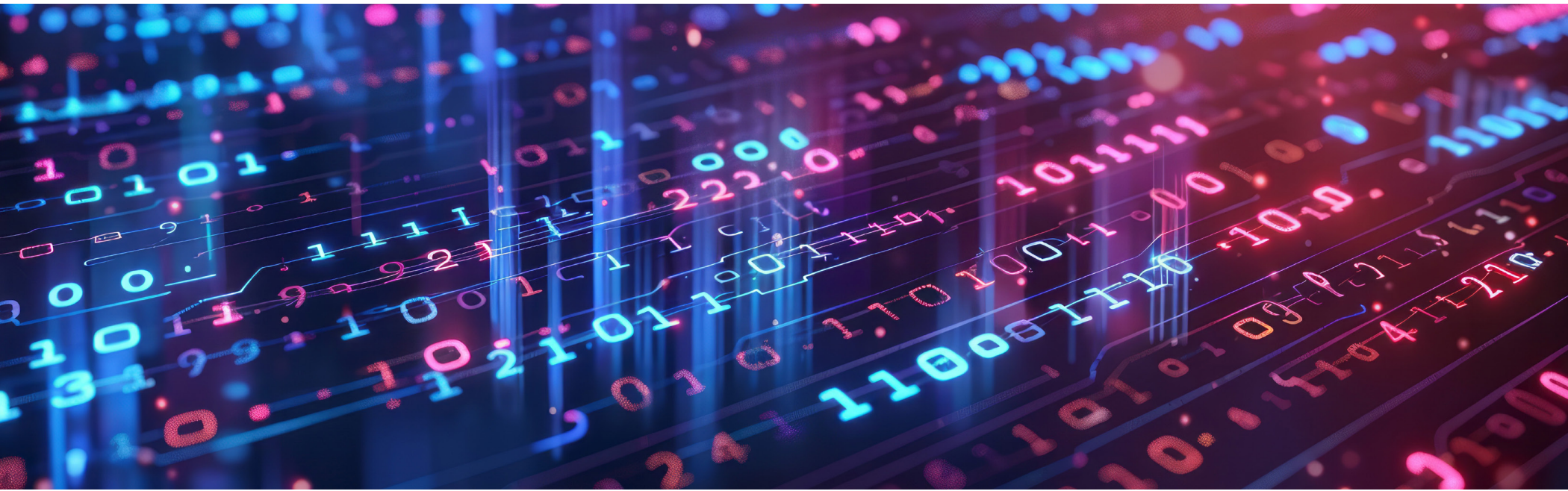

# Curated Standards and References

**[a]** The System Package Data Exchange® (SPDX®) Specification Version 3.0 - https://spdx.dev/use/specifications/

**[b]** IEEE 7000-2021 Model Process for Addressing Ethical Concerns during System Design is a best practices standard that provides a model process for identifying and addressing ethical concerns during system design. The information to be documented are system limitations, general information about AI systems, relevant stakeholder / intended use, identifying bias factors, explainability, runtime monitoring and accountability.

**[c]** IEEE 7001-20210 Transparency of Autonomous Systems is a best practice standard that provides guidelines for developing transparent AI systems. The information to be documented for compliance is: Information about application, limitations, and intended use, data used, algorithms employed and rationale for specific decision, explanations of the system's outputs and actions in a manner that is understandable to relevant stakeholders. Metrics (accuracy), bias / fairness, application's governance and who's accountable, systems security and privacy measures including any potential vulnerabilities and risks.

**[d]** IEEE 7002-2022 Data Privacy Processing is a best practices standard that provides guidelines for addressing bias in the design of autonomous and intelligent systems. The information to be documented is to identify potential sources of bias in the system, including data, algorithms, and human decision-making, potential impact of bias on the system's performance, fairness, and safety, mitigate or eliminate identified sources of bias. explain bias concerns to relevant stakeholders.

**[e]** IEEE 7005-2021 Transparent Employer Data Governance is a best practice standard which provides a process model to help identify and address ethical concerns in the design of autonomous and intelligent systems (A/IS). The documentation to be documented is all relevant Stakeholders (Section 6.2.1), any ethical concerns associated with the A/IS, such as issues related to privacy, fairness, transparency, and accountability.

**[f]** IEEE 7007-2021 Ethically Driven Robotics and Automation Systems is a best practice standard that provides guidance on the development of an ontological model that can be used to identify and address ethical concerns in the design, development, and deployment of robotics and automation systems. The details to be documented are to identify all relevant stakeholders who may be affected by the robotics and automation system (RAS), including direct and indirect users, and those who may be impacted by its operation.

**[g]** IEEE 7010-2020 - Impact of Autonomous and Intelligent Systems on Human Well-Being is a best practice standard. The assessment should log strategies to mitigate identified risks to human well-being. What safeguards, or developing policies and procedures to address the concerns. Evaluate Impacts and Mitigation Strategies and should evaluate the effectiveness of their impact mitigation strategies, and make any necessary adjustments.

**[h]** IEEE 7014-Ethical considerations in Emulated Empathy in Autonomous and Intelligent Systems is a standard that covers a range of design considerations for empathic A/IS, including user experience, accessibility, feedback mechanisms, and data privacy. It also provides guidance on conducting risk assessments to identify potential ethical risks associated with empathic A/IS, and developing strategies to mitigate those risks.

**[i]** IEEE 7009-2024 - IEEE Standard for Fail-Safe Design of Autonomous and Semi-Autonomous Systems provides a comprehensive framework for the fail-safe design of autonomous and semi-autonomous systems. This standard aims to ensure the safety, reliability, and robustness of such systems by addressing potential failure modes and implementing

strategies to mitigate risks. The standard covers various aspects of fail-safe design, including system architecture, fault detection and recovery mechanisms, and safety assurance processes.

**[j]** [ISO/IEC 5962:2021 Information technology SPDX® Specification V2.2.1](#) is a format for communicating information about software components and their relationships in a supply chain. The standard provides a common language and structure for describing software components, including their licenses, copyrights, security vulnerabilities, and other metadata. It also defines a set of best practices for creating and sharing SPDX documents, which can be used to facilitate the exchange of software component information between organizations and individuals.

**[k]** [ISO/IEC FDIS 5338 Information technology — AI — AI system life cycle processes](#) is focused on the life cycle processes of artificial intelligence (AI) systems. It provides a framework for the design, development, deployment, operation, maintenance, and retirement of AI systems, with the goal of promoting safe, effective, and ethical use of AI. AI system requirements: The developer should document the requirements for the AI system, including functional, performance, and safety requirements, as well as any ethical and societal considerations. The documented assessment includes the architecture, algorithms, and data used. The documentation should also include information on how the AI system addresses ethical and societal considerations, such as transparency, explainability, fairness, and robustness. should include information on how the AI system was tested, as well as any results or metrics. developers should document the deployment and operation of the AI system, including information on how the system is monitored, maintained, and updated, AI system decommissioning, should document any risk assessments conducted during the AI system life cycle, including assessments of ethical and societal risks. This should include information on how risks were identified, assessed, and mitigated, should document the mechanisms in place for accountability and responsibility in the design and development of the AI system, including clear lines of responsibility, documentation of design decisions, and processes for addressing ethical and societal concerns, any processes in place for continuous improvement of the AI system, including ongoing monitoring and evaluation of ethical and societal considerations, and mechanisms for feedback and learning and regulatory compliance.

**[l]** [ISO 13475 - Medical](#) Devices is an internationally agreed standard that sets out the requirements for a quality management system specific to the medical devices industry. It provides a framework for organizations involved in the design, production, installation, and delivery of medical devices to demonstrate their ability to provide medical devices and related services that consistently meet customer and regulatory requirements. The key features of the standard are: Regulatory Compliance that ensures that medical devices meet regulatory requirements, including those set by the FDA, European Union, and other global authorities, incorporates risk management activities during product realization to ensure the safety and effectiveness of medical devices. Helps organizations identify, analyze, evaluate, control, and monitor risks associated with medical devices.It rRequires organizations to maintain comprehensive documentation, including a quality manual, documented procedures, and records, ensures traceability and provides evidence of compliance and effectiveness. Ensures that the QMS is regularly reviewed and updated to maintain its effectiveness and using SPDX is a good solution analysis, and use of data to make informed decisions. ISO 13485:2016 is a critical standard for the medical devices industry, providing a robust framework for quality management that ensures the safety, effectiveness, and regulatory compliance of medical devices. By adhering to the requirements of ISO 13485, organizations can enhance their reputation, improve customer satisfaction, and gain access to global markets.

**[m]** [Code of Federal Regulations (CFR) Part 814 – Premarket Approval of Medical Devices](#) is a critical regulation for ensuring the safety and effectiveness of medical devices before they enter the market. By adhering to these requirements, manufacturers can demonstrate that their devices meet the FDA's rigorous standards, protecting public health and facilitating market access.

**[n]** [FDA Rules and Regulations](#) - The US FDA's rules and regulations are essential for maintaining public health and safety by ensuring that products meet high standards of quality, safety, and effectiveness. By providing comprehensive guidance, conducting inspections, and enforcing compliance, the FDA plays a critical role in protecting consumers and promoting innovation in the healthcare and food industries. The FDA's website offers a wealth of resources for understanding and complying with these regulations, making it a valuable tool for industry professionals, researchers, and the public. SPDX is recommended as a tool for product and AI inventory.

**[o]** Data, Analytics, and Artificial Intelligence Adoption Strategy. US Department of Defense. https://media.defense.gov/2023/Nov/02/2003333300/-1/-1/1/DOD_DATA_ANALYTICS_AI_ADOPTION_STRATEGY.PDF

**[p]** Mitchell, Margaret, Simone Wu, Andrew Zaldivar, Parker Barnes, Lucy Vasserman, Ben Hutchinson, Elena Spitzer, Inioluwa Deborah Raji, and Timnit Gebru. "Model cards for model reporting." In Proceedings of the conference on fairness, accountability, and transparency, pp. 220-229. 2019.

**[q]** Arnold, Matthew, Rachel KE Bellamy, Michael Hind, Stephanie Houde, Sameep Mehta, Aleksandra Mojsilović, Ravi Nair et al. "FactSheets: Increasing trust in AI services through supplier's declarations of conformity." IBM Journal of Research and Development 63, no. 4/5 (2019): 6-1.

**[r]** Gebru, Timnit, Jamie Morgenstern, Briana Vecchione, Jennifer Wortman Vaughan, Hanna Wallach, Hal Daumé Iii, and Kate Crawford. "Datasheets for datasets." Communications of the ACM 64, no. 12 (2021): 86-92.

**[s]** Hugging Face Model Card Template https://github.com/huggingface/huggingface_hub/blob/v0.24.6/src/huggingface_hub/templates/modelcard_template.md

**[t]** Build a Handwritten Text Recognition System using TensorFlow https://towardsdatascience.com/build-a-handwritten-text-recognition-system-using-tensorflow-2326a3487cd5

**[u]** IBM Face Sheet Introduction https://aifs360.res.ibm.com/introduction

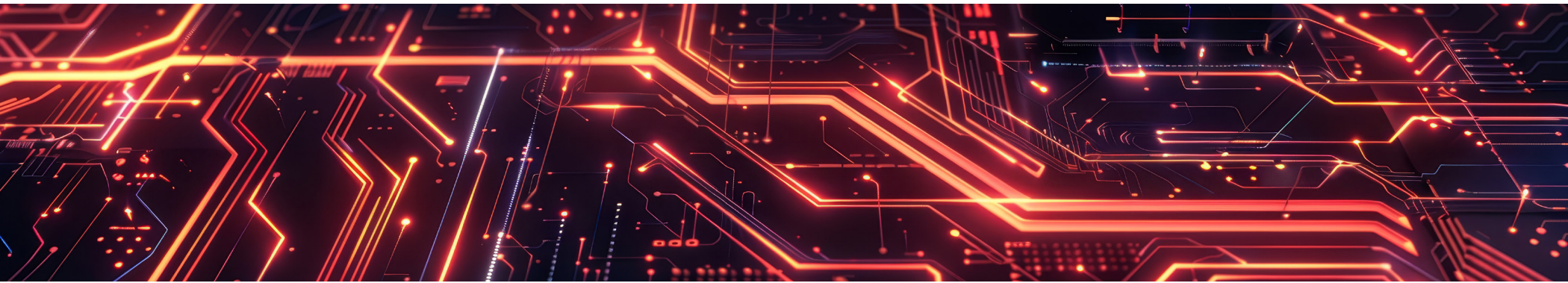

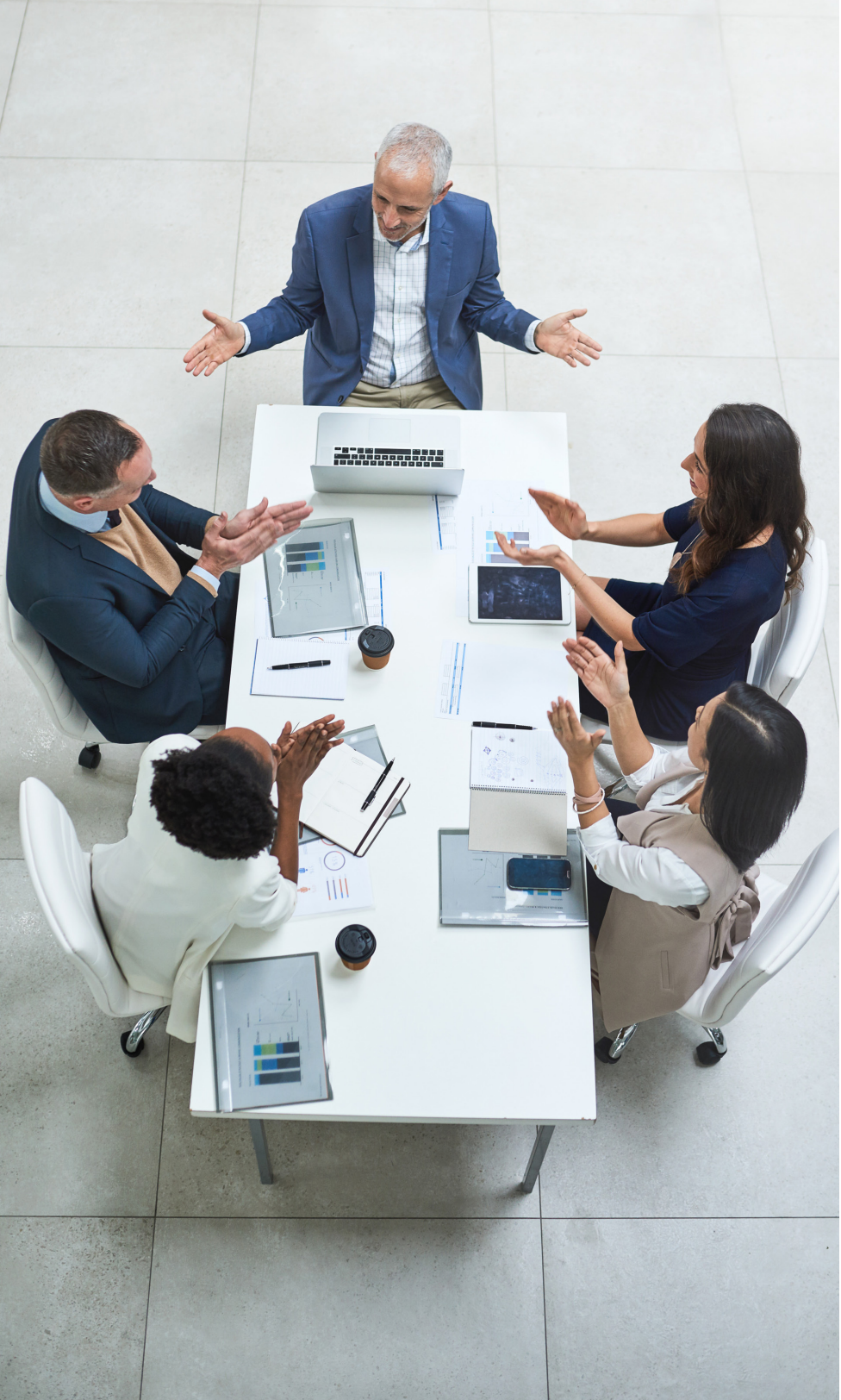

## Acknowledgments

The authors would like to thank the members of the AI & Dataset Profiles working group for their constructive input and helping us form a useful set of AI BOM fields. The authors would also like to thank the following reviewers for their comments on drafts of this paper:

- Alexios Zavras
- Gary O'Neall
- Matt White
- Michael Dolan
- Robert Martin

- Victor Liu
- Ibrahim Haddad
- Scott Bennet
- Michael Hind
- Steve Winslow

# About the Authors

**Karen Bennet**

Karen is an experienced senior engineering leader with more than 30 years in the software development business in both open and closed source solutions. She has successfully deployed large-scale AI platforms across self-driving cars, healthcare, financial, robotic and retail recommendation industries. She previously worked as a senior engineering leader at Red Hat, IBM, Yahoo and multiple startups in the AI platform/tools area. Karen is heavily involved with AI standardization as an AI expert with ISO, IEEE, Linux Foundation, NIST, CISA, EU AI Act, and Canada AI Act. Co-leads for the AI and Dataset Profiles, Karen holds 14 patents in the AI area, and has published technical papers on a variety of technical issues associated with AI.

**Gopi Krishnan Rajbahadur**

Gopi Krishnan Rajbahadur is a Senior Staff Researcher at Huawei's Centre for Software Excellence in Canada. He is currently working on software engineering for Large Language Models and the governance of AI datasets. His research interests encompass SE for AI, utilizing AI for SE, and the development of AI software that can be regulated. He is also an active contributor to the field of SE and AI standards, serving as the co-lead for the AI and Dataset Profiles in the ISO SPDX standard. He also co-founded the open-source initiative OpenDataology and frequently presents at Open Source Summits. His research has been featured in prominent SE publications such as TSE, TOSEM, EMSE, and ICSE.

**Arthit Suriyawongkul**

Arthit Suriyawongkul is a doctoral researcher at the Science Foundation Ireland Research Centre for AI-Driven Digital Content Technology (ADAPT) and the SFI Centre for Research Training in Digitally-Enhanced Reality at Trinity College Dublin. His system accountability ontology research focuses on the intersection of software engineering standards and legal frameworks, particularly the EU AI Act, to promote accountable AI development. Arthit actively contributes to the ISO submission of SPDX standard. Beyond his academic pursuits, Arthit co-founded and volunteers for Thai Netizen Network, a civil rights organization.

**Kate Stewart**

Kate Stewart is VP of Dependable Embedded Systems at the Linux Foundation. With over 30 years of experience in the software industry, she has held a variety of roles and worked as a developer in Canada, Australia, and the US and for the last 20 years has worked with open source software development teams around the world. Kate was one of the founders of SPDX, and is currently one of the technical leads for the project. She was also the co-lead for the NTIA SBOM formats and tooling working group, and has been co-leading the CISA SBOM tooling and implementation working group. Her current focus is helping open source projects being used in embedded markets to adopt best practices for safety, security and license compliance. Since joining The Linux Foundation in 2015, she has launched the Real Time LInux, Zephyr and ELISA Projects, among others.

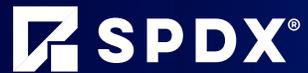

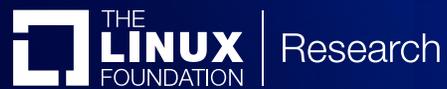

**About Linux Foundation Research**

Founded in 2021, Linux Foundation Research explores the growing scale of open source collaboration, providing insight into emerging technology trends, best practices, and the global impact of open source projects. Through leveraging project databases and networks, and a commitment to best practices in quantitative and qualitative methodologies, Linux Foundation Research is creating the go-to library for open source insights for the benefit of organizations the world over.

x.com/linuxfoundation

youtube.com/user/TheLinuxFoundation

facebook.com/TheLinuxFoundation

GITHUB.COM/LF-ENGINEERING

linkedin.com/company/the-linux-foundation